\documentclass[review,12pt]{elsarticle}
\usepackage{lineno,hyperref}
\usepackage{graphicx}
\usepackage{geometry}
\usepackage{caption}
\usepackage{subcaption}
\usepackage[linesnumbered,ruled,lined]{algorithm2e}
\usepackage{float}
\usepackage{tikz}
\usepackage{epstopdf}
\usepackage{amssymb}
\usepackage{amsmath}
\usepackage{indentfirst}
\usepackage{subcaption}
\usepackage{footnote}
\usepackage{array}
\usepackage{cases}
\usepackage{multirow}
\usepackage{ upgreek }
\usepackage{makecell}
\journal{CMAME}
\begin{document}
%
%
\begin{frontmatter}
		\title{A fourth-order kernel for improving numerical accuracy 
		and stability in Eulerian and total Lagrangian SPH}
		\author{Zhentong Wang }
		\ead{zhentong.wang@tum.de}
		\author{Bo Zhang}
		\ead{bo.zhang@tum.de}
		\author{Oskar J. Haidn}
		\ead{oskar.haidn@tum.de}
        \author{Xiangyu Hu \corref{mycorrespondingauthor}}
		\ead{xiangyu.hu@tum.de}
		\address{TUM School of Engineering and Design, 
		Technical University of Munich, Garching, 85747, Germany}
		\cortext[mycorrespondingauthor]{Corresponding author. }
\begin{abstract}
The error of smoothed particle hydrodynamics (SPH) using kernel for particle-based approximation
mainly comes from smoothing and integration errors.  
The choice of kernels has a significant impact on the numerical accuracy, stability and computational efficiency. 
At present, the most popular kernels such as B-spline, truncated Gaussian (for compact support), Wendland kernels have 2nd-order smoothing error 
and Wendland kernel becomes mainstream in SPH community as its stability and accuracy.
Due to the fact that the particle distribution after relaxation can achieve fast convergence of integration error respected to support radius, 
it is logical to choose kernels with higher-order smoothing error to improve the numerical accuracy.
In this paper, 
the error of 4th-order Laguerre-Wendland kernel proposed by Litvinov et al. \cite{litvinov2015towards} is revisited and  
another 4th-order truncated Laguerre-Gauss kernel is further analyzed and considered to replace the widely used Wendland kernel.
The proposed kernel has following three properties: 
One is that it avoids the pair-instability problem during the relaxation process, unlike the original truncated Gaussian kernel,  
and achieves much less relaxation residue than Wendland and Laguerre-Wendland kernels;
One is the truncated compact support size is the same as the non-truncated compact support of Wendland kernel, 
which leads to both kernels' computational efficiency at the same level;
Another is that the truncation error of this kernel is much less than that of Wendland kernel.
Furthermore,
a comprehensive set of $2D$ and $3D$ benchmark cases on Eulerian SPH for fluid dynamics 
and total Lagrangian SPH for solid dynamics validate the considerably improved numerical accuracy 
by using truncated Laguerre-Gauss kernel without introducing extra computational effort.

\end{abstract}

\end{frontmatter}
%
%
\section{Introduction}\label{referebce}
Smoothed particle hydrodynamics (SPH) is a meshless method 
originally proposed by Lucy \cite{lucy1977numerical}, Gingold and Monaghan \cite{monaghan1994simulating}
and has been widely used in fluid dynamics \cite{monaghan1994simulating}, 
solid mechanics \cite{libersky1993high}, and other scientific and industrial applications \cite{randles1996smoothed,longshaw2015automotive,gotoh2021entirely,sun2021generic}.
Since the particle-based approximation in SPH is formulated with 
a Gaussian-like smoothing kernel function with compact support \cite{gingold1977smoothed}, 
proper choice of the latter is crucial,
as already found and explored in Refs. \cite{liu2010smoothed, swegle1995smoothed,morris1996analysis},
for the numerical accuracy, stability and computational efficiency.
Generally, there are two factors determining the accuracy of SPH method,
i.e. smoothing and integration errors \cite{quinlan2006truncation, litvinov2015towards}.
While the leading vanishing moments of the kernel function define the smoothing error,
the particle summation on neighbor particles within the cut-off radius defines the integration error.
The current mainstream kernel functions used in SPH, such as B-spline, Wendland, 
truncated Gaussian (for compact support) and others,
are monotonic and give 2nd-order smoothing error as only the first moment vanishes. 
Although truncated Gaussian kernel is a nature choice for SPH,
it has not been widely used compare to other non-truncated, 
such as B-spline and Wendland, kernels.
One reason is that, in order to achieve sufficient small integration error, 
the truncated region  
is much larger than that of other non-truncated kernels
and results in lower computational efficiency.
The other reason is that, similarly to B-spline kernel, 
it causes pair-instability problem 
in which particles tend to appear pair clumping \cite{bui2021smoothed,swegle1995smoothed}.
In recent years, 
Wendland kernel has gained popularity in SPH community due to 
its moderate compact support for sufficient accuracy and 
ability to overcome pair-instability problem.

With given smoothing kernel, 
beside the size of compact support,
the integration error strongly relies on particle distribution \cite{quinlan2006truncation}.
Litvinov et al. \cite{litvinov2015towards} found 
that particles relaxed from random initial position and constant background pressure
is able to achieve the same accuracy as those located on uniform lattice positions.
Since relaxed particle distribution can be obtained for complex geometries, 
it is more applicable than uniform lattice distribution for practical applications.
Another finding in Ref. \cite{litvinov2015towards} is that, 
under relaxed particle distribution, 
the convergence rate of integration error (8th-order respected to support radius) 
can be much higher than that of smoothing error. 
Therefore, two non-monotonic, namely truncated Laguerre–Gauss and Laguerre–Wendland, 
kernels with 4th-order smoothing error has been considered 
for improving the overall accuracy of SPH approximations.
However, these kernels have not been applied for practical SPH algorithms,
due to the fact that, for moderate compact support, 
the actual integration error of Laguerre–Wendland kernel 
is much larger than that obtained by the original 2nd-order Wendland kernel.
Note that, the truncated Laguerre–Gauss kernel has not been tested in Ref. \cite{litvinov2015towards} 
probably due to the above-mentioned pair-instability problem of the original truncated Gaussian kernel.

In the present work, 
the large integration error of Laguerre–Wendland kernel in moderate compact support is revisited,
and the source of error has been analyzed according to the kernel profiles.
Based on this, the truncated Laguerre–Gauss kernel is chosen to improve the overall accuracy of SPH approximations.
Quite counter intuitively, further numerical tests show 
that not only the chosen kernel with moderate neighboring particles does not experience pair-instability problem like the original 2nd-order counterpart, 
but also able to achieve much less relaxation residue than that of Wendland kernel.
The chosen kernel has been applied to Eulerian and 
total Lagrangian SPH formulations for fluid and solid dynamics problems, respectively.
The numerical tests shown that considerable higher accuracy has been achieved compared with 
that of original Wendland kernel without introducing extra computational effort.

The structure of this paper is as follows: 
Section \ref{Kernel analysis} gives the preparations of error analysis and revisites Laguerre-Wendland kernel. 
Also, 
the errors of Laguerre-Wendland and truncated Laguerre-Gauss kernels are analyzed,
and the formulation as well as properties of the latter are discussed in detail.
Section \ref{Methodology} introduces standard Eulerian formulation with its extensions and total Lagrangian SPH formulas. 
The extensions include the incorporation of dissipation limiters to decrease numerical dissipation, 
the utilization of particle relaxation and kernel correction matrix to ensure zero-order and first-order consistency, respectively. 
Moving on to Section \ref{Numerical results}, 
a series of numerical examples are employed to demonstrate the performance and computational efficiency of the proposed kernel.
All computational codes utilized in this study have been made publicly available 
through the SPHinXsys repository \cite{zhang2021sphinxsys}, 
accessible at both https://www.sphinxsys.org and https://github.com/Xiangyu-Hu/SPHinXsys.
%
%
\section{Kernel analysis}\label{Kernel analysis}
%
%
\subsection{Preparations of error analysis}
The SPH approximation of gradient of a function field $ f(\mathbf{r})$ at a particle position $\mathbf{r}_{i}$ can be derived as 
following steps
\begin{equation} \label{Truncation error}
	\nabla f(\mathbf{r}_i) \approx \int \limits_V \nabla f(\mathbf{r}) W(\mathbf{r}_{i}-\mathbf{r}, h)dV
	=-\int \limits_V f(\mathbf{r}) \nabla W(\mathbf{r}_{i}-\mathbf{r}, h)dV\approx -\sum_{j} f(\mathbf{r}_j) \nabla W_{ij} V_{j},
\end{equation}
where $V_{j}$ is the volume of the neighboring particle $j$ respect to paricle $i$, 
 $ W(\mathbf{r}_{i}-\mathbf{r}, h)$ is a smooth kernel function with $h$ denoting the smoothing length and 
 the gradient of kernel function 
$\nabla W_{ij}=\nabla W(\mathbf{r}_{i}-\mathbf{r}_{j}, h)$ .
The first step is the smoothing approximation 
where the Dirac delta function is replaced by a smooth kernel function 
and introduces the error called smoothing error $E_s$.
The second step is integration by parts 
following the assumption that the kernel function is zero at the domain boundary.
The third step is the approximated integration by summation over all neighboring particles and introduces the error called integration error $E_r$.
Afterwards, the truncation error $E_t$ is given by $E_t=E_s+E_r$.
For simplicity of writing, we simplify $ f(\mathbf{r})$ at position $\mathbf{r}_{i}$ and $\mathbf{r}_{j}$ to $f_{i}$ and $f_{j}$ respectively 
in the following content.
Following Ref. \cite{zhang2022smoothed},
Eq. \eqref{Truncation error} can be modified in a strong form as 
\begin{equation}\label{strong form}
	\nabla f_{i} =\nabla f_{i} -f_{i}\nabla 1 \approx \sum_{j} f_{ij} \nabla W_{ij} V_{j},
\end{equation}
with $f_{ij}=f_{i}-f_{j}$. 
Also, 
the modification in a weak form as
\begin{equation}\label{weak form}
	\nabla f_{i}=\nabla f_{i}+f_{i}\nabla 1 \approx -2\sum_{j} \bar{f}_{ij} \nabla W_{ij} V_{j},
\end{equation}
with $\bar{f}_{ij}=(f_{i}+f_{j})/2$.
Due to the fact that the weak form is applied in the momentum conservation equations, 
we employ Eq. \eqref{weak form} to investigate the error estimation in the later section.
As is mentioned, 
the particle distribution plays a crucial role in influencing the integration error \cite{litvinov2015towards, yang2014new}. 
Therefore, 
to achieve a high-quality particle distribution in practical applications and ensure zero-order consistency, 
i.e. $\sum_{j} V_{j} \nabla W_{i j}=\mathbf{0}$, 
we employ particle relaxation \cite{zhu2021cad} before the error analysis 
and the acceleration in the process is calculated by 
\begin{equation}\label{relaxation acceleration}
    \mathbf{a}_{i}=-\frac{2}{\rho}_{i} \sum_{j}V_{j} \nabla W_{i j},
\end{equation}
with $m$ denoting the mass.
Note that the SPH approximation for the derivative of kernel function directly influence the flux calculation in conservation equations, 
we further introduce the kernel correction matrix \cite{randles1996smoothed} given by
\begin{equation}\label{kernel correction matrix}
	\mathbf{B}_i = -\left( \sum_{j} \mathbf r_{ij} \otimes \nabla W_{i j}  V_ {j}\right)^{-1}
\end{equation}
to compensate the error and satisfy the first-order consistency.
%
%
\subsection{Error analysis using Laguerre-Wendland kernel}\label{Error analysis using Laguerre-Wendland kernel}
As shown in Litvinov et al. \cite{litvinov2015towards},
Laguerre-Wendland and Laguerre-Gauss kernels have the second vanishing moment and 4th-order smoothing error.
Also,
the former had been studied in Litvinov et al. \cite{litvinov2015towards} but has not employed in practice 
due to its larger integration error than the widely used Wendland kernel.
Here, 
we revisit this kernel to explore the reason for the excessive integration error.
By comparing the profiles of the kernels and their derivatives, as shown in Figure \ref{kernel function and its derivative},  
we find that the gradient magnitude of Laguerre-Wendland kernel is much larger than that of Wendland kernel.
Note that the SPH approximation is highly relied on the size of compact support, 
which is usually adopted as $2h$ in practice where smoothing length $h=\kappa dp$ with $dp$ denoting the initial particle spacing and 
$\kappa$ determining the number of neighboring particles.
With these observations, 
it is straightforward to consider that 
the possible reason for Laguerre-Wendland kernel introducing larger integration error is that 
the number of neighboring particles is insufficient to resolve the large gradient magnitude within a moderate compact support. 
Based on this, 
we apply different $\kappa$ incluidng $\kappa=1.3$, $2.0$ and $2.5$ meaning different total number of neighboring particles 
for analysing the integration error.
Note that the corresponding resolutions are adopted according to the differnet $\kappa$ to keep smoothing length constant, 
i.e. the smoothing error is unchanged.
For the quantitative analysis, 
following Eq. \eqref{weak form},
$L_{1}$ and $L_\text{infinity}$ normalizations are given by
\begin{equation}\label{error normalizations}
\begin{cases} L_{1}(df/dx) = \frac{1}{N}\sum \limits_{i} \left|\nabla f_{i}-2\sum_{j} \bar{f}_{ij} \nabla W_{ij} V_{j} \right| \\ 
	L_{\text{infinity}}(df/dx)=\mathop{max}\limits_{i}\left|\nabla f_{i}-2\sum_{j} \bar{f}_{ij} \nabla W_{ij} V_{j} \right|  \end{cases}, 
\end{equation}
with $N$ and $df/dx$ denoting the total number of particles and the derivative of the function field, respectively.
Here, 
the error obtained in Eq. \eqref{error normalizations} is the truncation error. 
When a sufficient number of neighboring particles is present 
and the truncation error cannot be further reduced due to the negligible impact of the integration error, 
the dominant factor in the truncation error is the smoothing error. 
Conversely, 
when an insufficient number of neighboring particles is available, 
the integration error takes the lead in contributing to the overall truncation error.
\begin{table}[]
\caption{Truncation errors of the SPH approximation for $df/dx$ with $f=1$ and $f=\sin{(2\pi x)}$  
using Wendland and Laguerre-Wendland kernels with different $\kappa$ without the kernel corrrection matrix.}
\centering
\begin{tabular}{ccccccc}
\hline
\multirow{2}{*}{Truncation errors} & \multirow{2}{*}{$\kappa$}   & \multicolumn{2}{c}{$L_{1}(df/dx)$} &  & \multicolumn{2}{c}{$L_{\text{infinity}}(df/dx)$} \\ \cline{3-4} \cline{6-7} 
						&                      & $f=1$      & $f=\sin{(2\pi x)}$       &  & $f=1$      & $f=\sin{(2\pi x)}$       \\ \hline
Wendland                & \multirow{2}{*}{1.3} & 0.034         & 0.123     &  & 0.259         & 0.400     \\
Laguerre-Wendland       &                      & 0.070         & 0.673     &  & 1.880         & 3.506     \\ \hline
Wendland                & \multirow{2}{*}{2.0} & 0.014         & 0.073     &  & 0.027         & 0.150     \\
Laguerre-Wendland       &                      & 0.010         & 0.018     &  & 0.028         & 0.046     \\ \hline
Wendland                & \multirow{2}{*}{2.5} & 0.018         & 0.078     &  & 0.032         & 0.161     \\
Laguerre-Wendland       &                      & 0.005         & 0.009     &  & 0.017         & 0.030     \\ \hline
\end{tabular}
\label{Laguerre Wendland kernel analysis without correction}
\end{table}
\begin{table}[]
\caption{Truncation errors of the SPH approximation for $df/dx$ with $f=1$ and $f=\sin{(2\pi x)}$ 
using Wendland and Laguerre-Wendland kernels with different $\kappa$ with the kernel corrrection matrix.}
\centering
\begin{tabular}{ccccccc}
\hline
\multirow{2}{*}{Truncation errors} & \multirow{2}{*}{$\kappa$}   & \multicolumn{2}{c}{$L_{1}(df/dx)$} &  & \multicolumn{2}{c}{$L_{\text{infinity}}(df/dx)$} \\ \cline{3-4} \cline{6-7} 
						&                      & $f=1$      & $f=\sin{(2\pi x)}$       &  & $f=1$      & $f=\sin{(2\pi x)}$       \\ \hline
Wendland                & \multirow{2}{*}{1.3} & 0.039         & 0.052     &  & 0.209         & 0.187     \\
Laguerre-Wendland       &                      & 0.077         & 0.067     &  & 2.334         & 2.319     \\ \hline
Wendland                & \multirow{2}{*}{2.0} & 0.014         & 0.042     &  & 0.026         & 0.067     \\
Laguerre-Wendland       &                      & 0.011         & 0.007     &  & 0.029         & 0.028     \\ \hline
Wendland                & \multirow{2}{*}{2.5} & 0.018         & 0.043     &  & 0.033         & 0.068     \\
Laguerre-Wendland       &                      & 0.005         & 0.003     &  & 0.017         & 0.015     \\ \hline
\end{tabular}
\label{Laguerre Wendland kernel analysis with correction}
\end{table}

In the study, 
a circular geometry of a diameter $D = 2$ with the resolution $dp = 1/60$ is applied and 
we test the SPH approximation for the derivatives of an constant function $f=1$ and a trigonometric function $f=\sin{(2\pi x)}$ 
using Wendland and Laguerre-Wendland kernels without and with the kernel correction matrix 
in this domain with the respective particle distributions obtained by their own kernels shown in Figure \ref{acceleration magnitude}, 
and Tables \ref{Laguerre Wendland kernel analysis without correction} and \ref{Laguerre Wendland kernel analysis with correction} 
list the truncation errors with different $\kappa$. 
Note that the boundary particles are not considered in the kernel
analysis due to the fact that these particles are given the assigned values and do not
evaluate the kernel gradient.
From the results, 
the truncation errors using Laguerre-Wendland kernel exhibit similar behavior to those obtained with Wendland kernel, 
declining with $\kappa \leq 2$ and remaining relatively unchanged at $\kappa=2.5$.
These results suggest that, 
while the smoothing error dominates at $\kappa \geq 2$, 
the integration error takes the lead at $\kappa=1.3$. 
For the condition of smoothing error dominant with  $\kappa \geq 2$, 
it is implied that the smoothing error of Laguerre-Wendland kernel is less than Wendland kernel, 
which agrees with the property of kernels' smoothing order.
In addition, 
given the reasonable truncation errors, 
while the impact of the kernel correction matrix on reducing the truncation error with the derivative of $f=1$ is not as apparent, 
this matrix can further minimize the error with that of $f=\sin{(2\pi x)}$ effectively, 
suggesting that the kernel correction matrix has distinct effects on higher-order terms.
%
%
\subsection{Truncated Lagueree-Gauss kernel}\label{Truncated Laguree-Gauss analysis}
As another 4th-order kernel, as shown in Figure \ref{kernel function and its derivative},
Laguerre-Gauss kernel with smaller gradient magnitude than Laguerre-Wendland kernel can be seen as a potential option 
for not introducing excessive integration error.
%
%
\subsubsection{Fundamentals of truncated Lagueree-Gauss kernel}\label{Truncated Laguree-Gauss kernel basics}
\begin{figure} 
	\centering
	\begin{subfigure}[b]{0.49\textwidth}
		\centering
		\includegraphics[trim = 0cm 0cm 0cm 0cm, clip, width=0.9\textwidth]{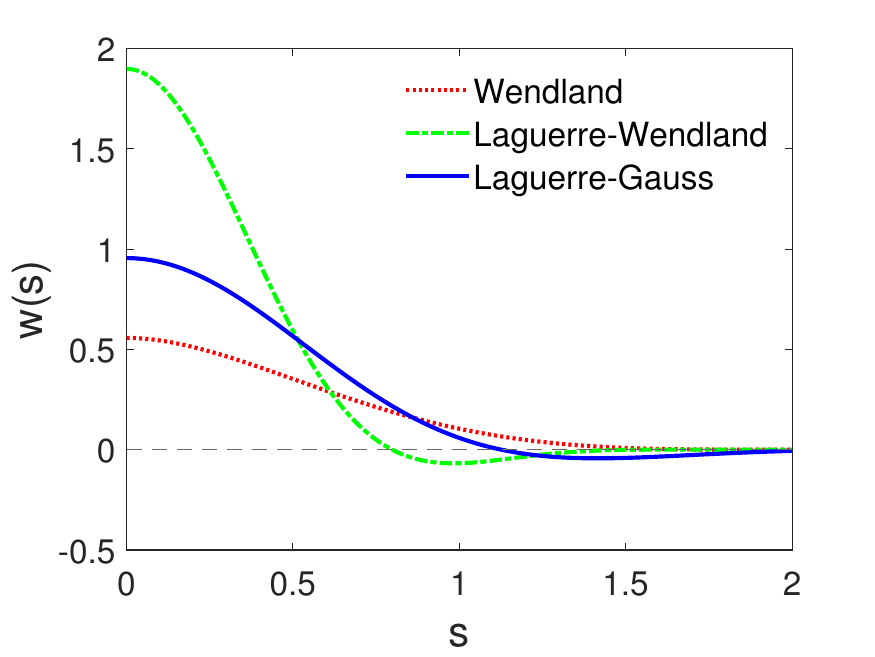}
	\end{subfigure}
	\begin{subfigure}[b]{0.49\textwidth}
		\centering
		\includegraphics[trim = 0cm 0cm 0cm 0cm, clip, width=0.9\textwidth]{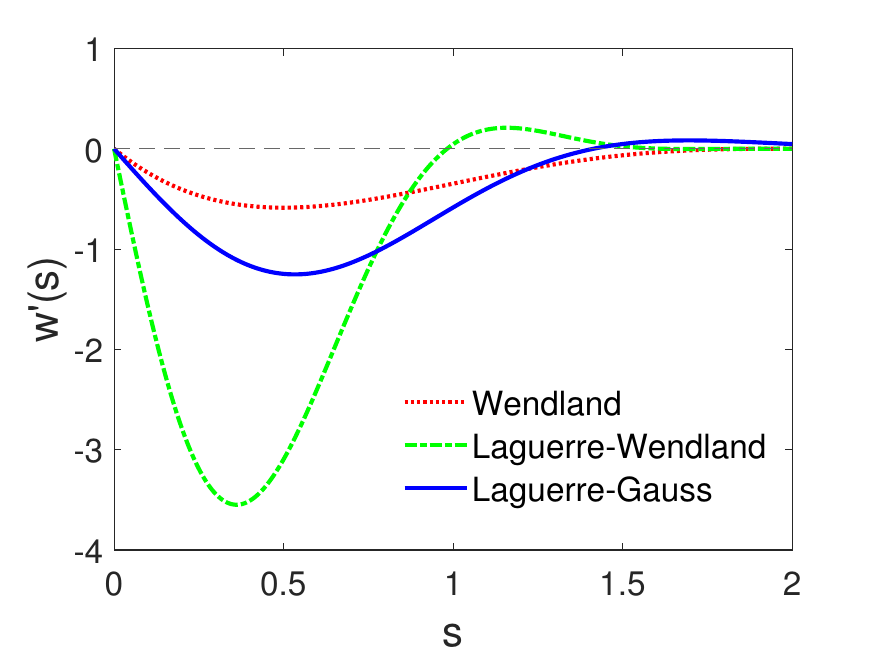}
	\end{subfigure}
	\caption{Laguerre-Gauss kernel and the comparison with Wendland as well as Laguerre-Wendland kernels with compact support as $2h$
	: kernel functions (left panel) and their derivatives (right panel).}
        \label{kernel function and its derivative}
\end{figure}

The expression of Laguerre-Gauss kernel \cite{litvinov2015towards} is given by
\begin{equation}\label{Laguerre-Gauss kernel}
	w(s)= \pi^{-1}L_{1}^{(1)}(s^{2})e^{-s^{2}},
	\end{equation}
where $L_{1}^{(1)}$ denotes the generalized Laguerre polynomial with non-negative Fourier transform and vanishing second-order moment  
as well as $s=\left|\mathbf{r} \right|/h$. 
For implementing the kernel in practice, we truncate compact domain to $2h$ 
and give the expression for truncated Laguerre-Gauss kernel as
\begin{equation}
w(s)=\alpha_{d}\begin{cases} (1-\frac{s^{2}}{2}+\frac{s^{4}}{6})e^{-s^{2}} & \text { if } 0 \leq s \leq 2 \\ 0 & \text { if } s>2  \end{cases}, 
\end{equation}
where $\alpha_{d}$ is the normalized coefficient in $d$ dimensional space with the value of $\alpha_{1}=8/(5\times\sqrt{\pi})$, 
$\alpha_{2}=3/\pi$ and $\alpha_{3}=8/\pi^{3/2}$. 
Also, 
the truncated Laguerre-Gauss kernel and its derivative and the comparisons with Wendland as well as Laguerre-Wendland kernels are presented in 
Figure \ref{kernel function and its derivative} showing that 
its gradient magnitude is larger than that of Wendland kernel but smaller than that of Laguerre-Wendland kernel 
and the value of kernel function at $2h$ is close to $0$. 
In numerical simulations, 
to avoid adding extra computational effort, 
we apply the compact support as $2h$ with smoothing length $h=1.3 dp$, 
meaning that the computational efficiency using truncated Laguerre-Gauss kernel is the same as that using Wendland kernel.
%
%
%
\subsubsection{Performance in stability}\label{Performance in stability}
To exploit the properties of truncated Laguerre-Gauss kernel, 
we firstly investigate and compare the relaxed particle distribution 
using Wendland, Laguerre-Wendland and truncated Laguerre-Gauss kernels in the same circular domain mentioned above.
\begin{figure}
	\centering
	\begin{subfigure}[b]{0.49\textwidth}
		\centering
		\includegraphics[trim = 0cm 0cm 0cm 0cm, clip, width=0.9\textwidth]{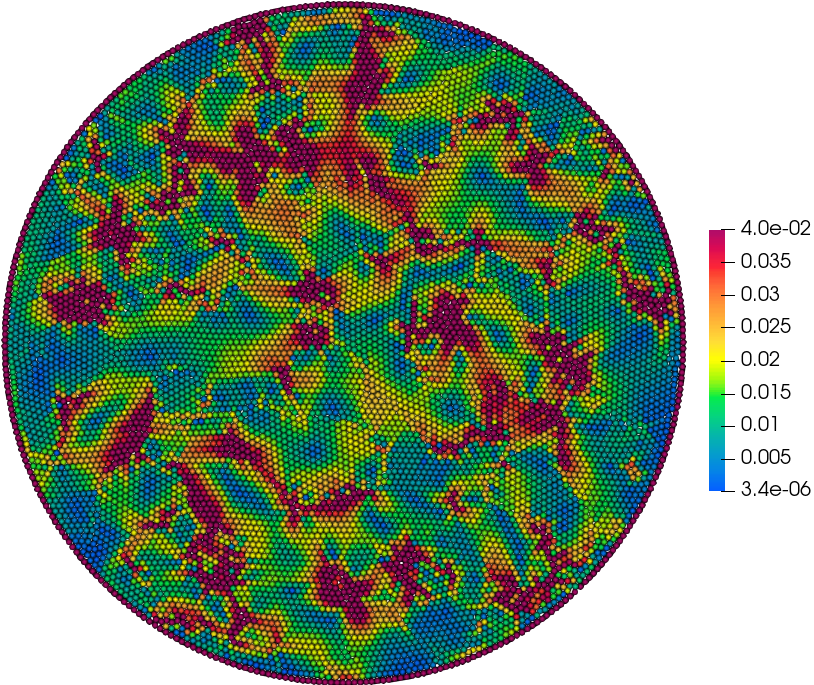}
	\end{subfigure}
		\centering
	\begin{subfigure}[b]{0.49\textwidth}
		\centering
		\includegraphics[trim = 0cm 0cm 0cm 0cm, clip, width=0.9\textwidth]{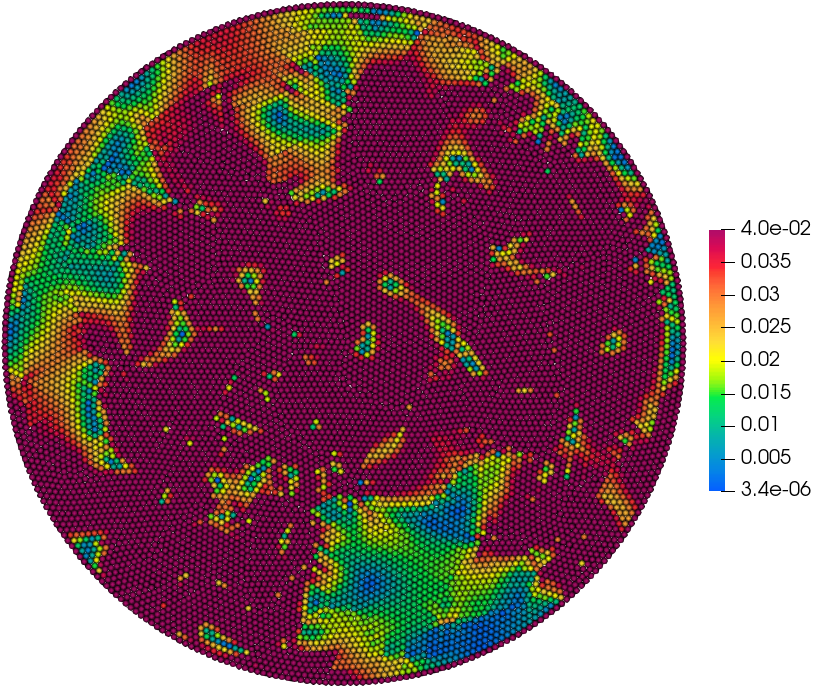}
	\end{subfigure}
		\centering
	\begin{subfigure}[b]{0.49\textwidth}
		\centering
		\includegraphics[trim = 0cm 0cm 0cm 0cm, clip, width=0.9\textwidth]{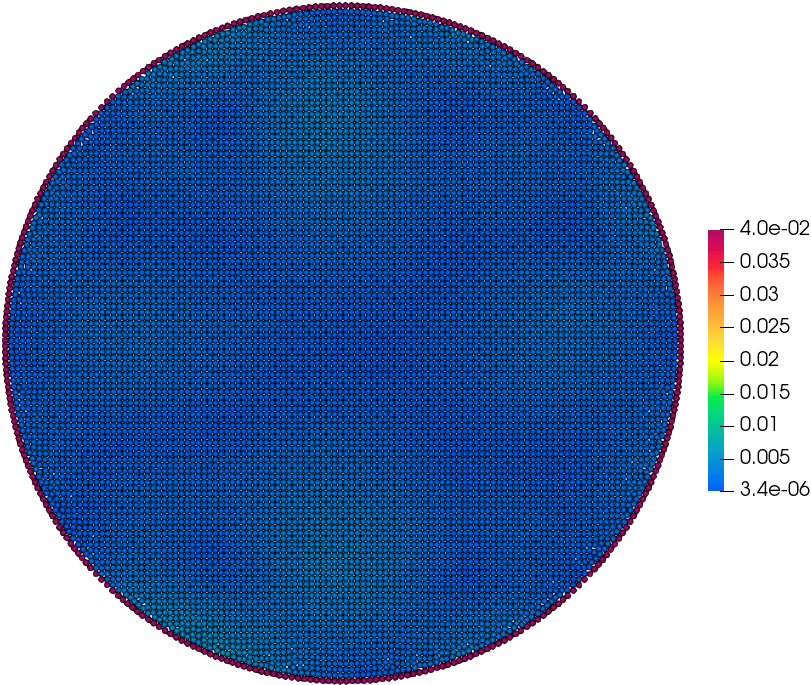}
	\end{subfigure}
	\caption{Particle relaxation residue ranging from $3.4\times 10^{-6}$ to $4.0\times 10^{-2}$ 
	using Wendland (top left panel), Laguerre-Wendland (top right panel) and truncated Laguerre-Gauss (bottom panel) kernels
	at the final relaxation iteration with the spatial resolution $dp=1/60$. 
	In the case, 
	the boundary particles are not considered due to its not relying on kernel approximation.
	Based on this, 
	the average relaxation residue and the variation between the maximum and minimum relaxation residues using truncated Laguerre-Gauss kernel 
	are approximately $0.004$ and $0.011$, respectively, 
	which are approximately one-tenth and one-twentieth of the values using Wendland kernel.
	Meanwhile, 
	the average relaxation residue of $0.07$ when employing Laguerre-Wendland kernel is approximately twenty times larger 
	than that of using truncated Laguerre-Gauss kernel 
	and the maximum relaxation residue of $1.88$ appears to be unreasonable.}
	\label{acceleration magnitude}
\end{figure}
Figure \ref{acceleration magnitude} shows the relaxation residue ranging from $3.4\times 10^{-6}$ to $4.0\times 10^{-2}$ using Wendland, 
Laguerre-Wendland and truncated Laguerre-Gauss kernels at the final relaxation iteration. 
Completely contrary to our intuition that 
truncated Laguerre-Gauss kernel is similar to original truncated Gaussian kernel in that it suffers from pair-instability problem with $\kappa=1.3$,
it can be observed that the particle relaxation using truncated Laguerre-Gauss kernel does not appear this problem. 
Similarly with the kernel analysis in section \ref{Error analysis using Laguerre-Wendland kernel}, 
the particles in the boundary are also omitted in the following analysis.
Based on this, 
truncated Laguerre-Gauss kernel yields an average relaxation residue of $0.004$ and a variation between the maximum and minimum relaxation residues of $0.011$ 
and these values represent roughly one-tenth and one-twentieth of those obtained using Wendland kernel.
Besides, 
Laguerre-Wendland kernel results in an average relaxation residue of $0.07$, 
which is approximately twenty times larger than that observed with truncated Laguerre-Gauss kernel. 
However, 
the maximum relaxation residue of $1.880$ using Laguerre-Wendland kernel is unacceptable and unreasonable. 
Therefore, 
truncated Laguerre-Gauss kernel ensures the numerical stability and obtains higher-quality particle distribution.
%
%
\subsubsection{Performance in accuracy}\label{Performance in accuracy}
Similarly with the analysis of Laguerre-Wendland kernel, 
we test the SPH approximation for the derivatives of same functions using truncated Laguerre-Gauss kernel to further verify the approximation accuracy.
Specifically, 
the truncation error using the derivitive of function $f=1$ is equal to the relaxation residue in Eq. \eqref{relaxation acceleration}.
\begin{table}[]
\caption{Truncation errors of the SPH approximation for $df/dx$ with $f=1$ and $f=\sin{(2\pi x)}$ 
using truncated Laguerre-Gauss kernel with different $\kappa$ without the kernel corrrection matrix.}
\centering
\begin{tabular}{ccccccc}
\hline
\multirow{2}{*}{Truncation errors}   & \multirow{2}{*}{$\kappa$} & \multicolumn{2}{c}{$L_{1}(df/dx)$} &  & \multicolumn{2}{c}{$L_{\text{infinity}}(df/dx)$} \\ \cline{3-4} \cline{6-7} 
									&                    & $f=1$      & $f=\sin{(2\pi x)}$       &  & $f=1$      & $f=\sin{(2\pi x)}$       \\ \hline
\multirow{3}{*}{Laguerre-Gauss} &       1.1               & 0.021         & 0.259     &  & 0.614         & 1.253     \\
								&       1.3               & 0.004         & 0.123     &  & 0.011         & 0.204     \\
								&       1.6               & 0.007         & 0.210     &  & 0.822         & 0.817     \\ \hline
\end{tabular}
\label{Laguerre Gauss kernel analysis without correction}
\end{table}
\begin{table}[]
\caption{Truncation errors of the SPH approximation for $df/dx$ with $f=1$ and $f=\sin{(2\pi x)}$ 
using truncated Laguerre-Gauss kernel with different $\kappa$ with the kernel corrrection matrix.}
\centering
\begin{tabular}{ccccccc}
\hline
\multirow{2}{*}{Truncation errors}   & \multirow{2}{*}{$\kappa$} & \multicolumn{2}{c}{$L_{1}(df/dx)$} &  & \multicolumn{2}{c}{$L_{\text{infinity}}(df/dx)$} \\ \cline{3-4} \cline{6-7} 
									&                    & $f=1$      & $f=\sin{(2\pi x)}$       &  & $f=1$      & $f=\sin{(2\pi x)}$       \\ \hline
\multirow{3}{*}{Laguerre-Gauss} &       1.1               & 0.021         & 0.061     &  & 0.603         & 0.448     \\
								&       1.3               & 0.004         & 0.004     &  & 0.011         & 0.009     \\
								&       1.6               & 0.006         & 0.024     &  & 0.784         & 0.364     \\ \hline
\end{tabular}
\label{Laguerre Gauss kernel analysis with correction}
\end{table}

Tables \ref{Laguerre Gauss kernel analysis without correction} and \ref{Laguerre Gauss kernel analysis with correction} 
list the truncation errors using truncated Laguerre-Gauss kernel. 
From the results, 
the truncation error in the case of insufficient number of neighboring particles with $\kappa=1.1$ is still relatively large.
Besides, 
the large errors obtained with $\kappa=1.6$ is due to the pair instability when the number of neighboring particles is too large.
Note that $\kappa=1.3$ is applied in practice as mentioned above to avoid extra computational effort. 
In the case of $\kappa=1.3$ with the derivative of latter function, 
we observe that the truncation errors using truncated Laguerre-Gauss kernel in Table \ref{Laguerre Gauss kernel analysis without correction} 
and Wendland kernel in Table \ref{Laguerre Wendland kernel analysis without correction} are comparable without the utilization of the kernel correction matrix.
However, 
the truncation error using the proposed kernel in Table \ref{Laguerre Gauss kernel analysis with correction} 
is significantly smaller than the errors using Wendland and Laguerre-Wendland kernels 
in Table \ref{Laguerre Wendland kernel analysis with correction},  
demonstrating that the former kernel cooperation with the kernel correction matrix offers much higher accuracy. 
This finding suggests the integration error using truncated Laguerre-Gauss kernel is lower compared to the other two kernels as 
the integration errors are dominant in using Wendland and Laguerre-Wendland kernels with $\kappa=1.3$.
Furthermore, 
the proposed kernel exhibits superior performance with respect to the average relaxation residue and the variation of the maximum and minimum residue. 
Specifically, 
in Tables \ref{Laguerre Gauss kernel analysis without correction}, 
the average relaxation residue and the variation are $0.004$ and $0.011$, respectively, 
which are approximately one-tenth and one-twentieth of the values 
listed in Tables \ref{Laguerre Wendland kernel analysis without correction} for the Wendland kernel,   
proving that truncated Laguerre-Gauss kernel enables obtain much less average relaxation residue and the variation 
than those using the other two kernels and these results agree with the particle distribution shown in Figure \ref{acceleration magnitude}.
%
%
\section{Governing equations and SPH methods}\label{Methodology}
Here, 
we consider an extended Eulerian SPH method for fluid dynamics \cite{monaghan1994simulating} 
and the standard total Lagrangian SPH method for solid mechanics \cite{libersky1993high,wu2023essentially}. 
%
%
\subsection{Extended Eulerian SPH}\label{Standard Eulerian SPH framework}
%
%
\subsubsection{Governing equatioins}\label{Eulerian SPH governing equatioins}
The conservation equations in the Eulerian framework can be described by
\begin{equation}\label{eqs:conservation}
\frac{\partial \boldsymbol{U}}{\partial t}+\nabla \cdot \boldsymbol F(\boldsymbol{U})=0,
\end{equation}
where $\boldsymbol{U}$ is the vector of the conserved variables, $\boldsymbol{F}(\boldsymbol{U})$ the corresponding fluxes. 
Here, they can be expressed specifically as
\begin{equation}
\label{eqs:flux}
\mathbf{U}=\left[\begin{array}{c}
\rho \\
\rho u \\
\rho v \\
\rho w \\
E
\end{array}\right], \quad \mathbf{F}=\left[\begin{array}{c}
\rho u \\
\rho u^{2}+p \\
\rho u v \\
\rho u w \\
u(E+p)
\end{array}\right]+\left[\begin{array}{c}
\rho v \\
\rho vu \\
\rho v^{2}+p \\
\rho vw \\
v(E+p)
\end{array}\right]+\left[\begin{array}{c}
\rho w \\
\rho wu \\
\rho wv \\
\rho w^{2}+p \\
w(E+p)
\end{array}\right],
\end{equation}
where $u$, $v$ and $w$ are the components of velocity, $\rho$ and $p$ denote the density and pressure, respectively. 
Here, $E=\frac{\rho{\mathbf{v}}^2}{2}+\rho e$ is the total energy per volume with $e$ the internal energy. 
The equation of state (EOS) is added to close the Eq. \eqref{eqs:conservation} by
\begin{equation}
p=\begin{cases} \rho(\gamma-1)e & \text { For compressible flows } \\ c^2(\rho-\rho_0) & \text { For weakly-compressible flows } \end{cases}, 
\end{equation}
where $\gamma$ is the heat capacity ratio, $\rho_{0}$ the reference density. 
Following the ideal gas equation in compressible flows and the weakly-compressible assumption in incompressible flows, 
the speed of sound is derived by
\begin{equation}
c=\begin{cases} \sqrt{\frac{\gamma p}{\rho}} & \text { For compressible flows } \\ 10U_{max} & \text { For weakly-compressible flows } \end{cases}.
\end{equation}
Here, $U_{max}$ the maximum velocity in the flow field to control the density variation less than $1\%$. 
Note that the energy equation is turned off in weakly-compressible flows. 
In the viscous weakly-compressible flows, we add a viscous force given by
\begin{equation}\label{viscous term}
	\left(\frac{\partial \mathbf{v}}{\partial t}\right)^{\left(\upsilon\right)}=\eta {\nabla}^{2}\mathbf{v}
\end{equation}
to the momentum equation.
%
%
\subsubsection{Standard Eulerian SPH discretization}\label{Eulerian SPH discretization}
Following Ref. \cite{vila1999particle},
the discretization form of the Eq. \eqref{eqs:conservation} can be written as
\begin{equation}\label{eqs:conservation-discretize}
\left\{\begin{array}{l}
\frac{\partial}{\partial t}\left(w_{i}\rho_{i}\right)+2 w_{i}\sum_{j} w_{j}  (\rho \mathbf{v})^{*}_{E, i j} \cdot \nabla W_{i j}=0 \\
\frac{\partial}{\partial t}\left(w_{i}\rho_{i} \mathbf{v}_{i}\right)+
2 w_{i}\sum_{j} w_{j} \left[(\rho \mathbf{v} \otimes \mathbf{v})^{*}_{E, i j}+p^{*}_{E, i j}\mathbb{I}\right] \cdot \nabla W_{i j}=0 \\
\frac{\partial}{\partial t}\left(w_{i}E_{i}\right)+ 2 w_{i}\sum_{j} w_{j} \left[(E\mathbf{v})^{*}_{E, i j}+
(p \mathbf{v})^{*}_{E, i j}\right] \cdot \nabla W_{i j}=0
\end{array},\right.
\end{equation}
where $w$ represents the volume of particle, 
$\mathbf{v}$ the velocity, $\mathbb{I}$ the identity matrix and terms $()^{*}_{E, i j}$ are solution of the Riemann problem \cite{vila1999particle}.
Here, 
$\nabla W_{i j}=\frac{\partial W_{i j}}{\partial r_{ij}}\mathbf{e}_{ij}$ denotes kernel gradient 
where $\mathbf{e}_{ij}=-\mathbf{r}_{ij}/r_{ij}$ with $\mathbf{r}_{ij}$ the displacement pointing from particle $j$ to $i$. 

To obtain the solution of the Riemann problem, 
three waves with the smallest speed $S_{l}$, middle speed $S_{\ast}$ and largest speed $S_{r}$ are utilized. 
Note that the middle wave distinguishes the two intermediate states as $(\rho_l^{\ast}, u_l^{\ast},p_l^{\ast})$ and $(\rho_r^{\ast}, u_r^{\ast},p_r^{\ast})$. 
For compressible flows, 
we employ the HLLC Riemann solver \cite{toro1994restoration,toro2019hllc} because of its ability in capturing the shock discontinuity, 
with the three wave speeds estimated as
\begin{equation}\label{three wave estimation}
\left\{\begin{array}{l}
S_{l}=u_{l}-c_{l} \\
S_{r}=u_{r}+c_{r}\\
S_{\ast}=\frac{\rho_{r} u_{r}\left(S_{r}-u_{r}\right)+\rho_{l} u_{l}\left(u_{l}-S_{l}\right)+p_{l}-p_{r}}{\rho_{r}\left(S_{r}-u_{r}\right)+
\rho_{l}\left(u_{l}-S_{l}\right)}
\end{array}.\right.
\end{equation}
Then, the intermediate states can be calucated as
\begin{equation}\label{conserved variables in HLLC}
\left\{\begin{array}{l}
p^{*}=p_{l}+\rho_{l}\left(u_{l}-S_{l}\right)\left(u_{l}-u^{*}\right)=p_{r}+\rho_{r}\left(S_{r}-u_{r}\right)\left(u^{*}-u_{r}\right) \\
\mathbf{v}_{l/r}^{*}=u^{*}\mathbf{e}_{ij}+ \left[\frac{1}{2}(\mathbf{v}_{l}+\mathbf{v}_{r})- \frac{1}{2}({u}_{l}+{u}_{r})\mathbf{e}_{ij}\right]\\
\rho_{l/r}^{*}=\rho_{l/r} \frac{\left(S_{l/r}-q_{l/r}\right)}{\left(S_{l/r}-u^{*}\right)}\\
E_{l/r}^{*}=\frac{\left(S_{l/r}-q_{l/r}\right) E_{l/r}-p_{l/r} q_{l/r}+p^{*} u^{*}}{S_{l/r}-u^{*}}
\end{array},\right.
\end{equation}
where $q$ represents the velocity magnitude and ${u}^{*}={S}_{\ast}$. 
In weakly-compressible flows, 
the intermediate states follows the assumption satisfying ${p}^{*}_{l}={p}^{*}_{r}={p}^{*}$ and ${u}^{*}_{l}={u}^{*}_{r}={u}^{*}$ 
and then a linearised Riemann solver can be expressed as \cite{toro2013riemann,wang2023eulerian}
\begin{equation}\label{linearised Riemann solver}
\left\{\begin{array}{l}
u^{*}=\frac{u_{l}+u_{r}}{2}+\frac{1}{2} \frac{\left(p_{L}-p_{R}\right)}{\bar{\rho} \bar{c}} \\
p^{*}=\frac{p_{l}+p_{r}}{2}+\frac{1}{2} \bar{\rho}  \bar{c} \left(u_{l}-u_{r}\right)
\end{array},\right.
\end{equation}
with $\bar{\rho}$ and $\bar{c}$ interface-particle averages.
%
%
\subsubsection{Eulerian SPH extensions}\label{Eulerian SPH extensions}
In this section, 
we implement the particle relaxation following Eq. \eqref{relaxation acceleration} and 
kernel correction matrix following Eq. \eqref{kernel correction matrix} and detail the dissipation limiters, 
which are used to improve stability and accuracy of Eulerian SPH method.
To maintain the conservation property, 
the corrected kernel gradient can be re-evaluated as 
\begin{equation}
	{\nabla}^{'} W_{i j}=\frac{\mathbf{B}_i+\mathbf{B}_j}{2} \nabla W_{i j}
\end{equation}
to replace the original kernel gradient $\nabla W_{i j}$ in Eq. \eqref{eqs:conservation-discretize}.
 
In order to reduce the numerical dissipation introduced by the Riemann problem, 
dissipation limiters \cite{wang2023eulerian,zhang2017weakly} are implemented and can greatly improve numerical accuracy. 
In the HLLC Riemann solver, the middle wave speed and pressure in Eq. \eqref{conserved variables in HLLC} with the limter can be rewritten as 
\begin{equation}
\left\{\begin{array}{l}
u^{*}=\frac{\rho_{l}u_{l}c_{l}+\rho_{r}u_{r}c_{r}}{\rho_{l}c_{l}+\rho_{r}c_{r}}+\frac{p_{l}-p_{r}}{\rho_{l}c_{l}+\rho_{r}c_{r}}\beta^{2}_{HLLC}\\
p^{*}=\frac{p_{l}+p_{r}}{2} +\frac{1}{2}\beta_{HLLC}\left[\rho_{r}c_{r}\left(u^{*}-u_{r}\right)-\rho_{l}c_{l}\left(u_{l}-u^{*}\right)\right] 
\end{array},\right.
\end{equation}
with the limiter as 
\begin{equation}\label{HLLC limiter}
\beta_{HLLC}=\min \left(\upeta_{HLLC} \max (\frac{u_{l}-u_{r}}{\bar{c}}, 0) , 1\right),
\end{equation}
where $\upeta_{HLLC}=1.0$ according the numerical tests.
Also, the linearised Riemann solver in Eq. \eqref{linearised Riemann solver} with the limiter can be rewritten as 
\begin{equation}\label{acoustic Riemann solver}
\left\{\begin{array}{l}
u^{*}=\frac{u_{l}+u_{r}}{2}+\frac{1}{2} \beta^{2}_{linearisd} \frac{\left(p_{L}-p_{R}\right)}{\bar{\rho} \bar{c}} \\
p^{*}=\frac{p_{l}+p_{r}}{2}+\frac{1}{2} \beta_{linearisd} \bar{\rho}  \bar{c} \left(u_{l}-u_{r}\right)
\end{array},\right.
\end{equation}
with the dissipation limiter $\beta_{linearisd}$ as
\begin{equation}\label{acoustic limiter}
\beta_{linearisd}=\min \left(\upeta_{linearisd} \max (\frac{u_{l}-u_{r}}{\bar{c}}, 0), 1\right).
\end{equation}
where $\upeta_{linearisd}=15.0$.
%
%
\subsection{Total Lagrangian SPH}\label{Standard total Lagrangian SPH framework}
In total Lagrangian SPH framework for elastic solid dynamics, the mass and momentum conservation equations can be expressed as 
\begin{equation}\label{total Lagrangian SPH governing equatioins}
\left\{\begin{array}{l}
\rho=J^{-1} \rho^0 \\
\rho^0 \frac{d\boldsymbol{v}_{i}}{dt}=\nabla^0 \cdot \mathbb{P}^{\mathrm{T}}
\end{array},\right.
\end{equation}
with $\rho^0$ and $\rho$ the initial and current density, respectively. 
Here, $J=det(\mathbb{F})$ with $\mathbb{F}$ denoting the deformation gradient tensor, $\boldsymbol{v}$ represents the velocity, 
and $\mathbb{P}$ as well as $T$ are the first Piola-Kirchhoff stress tensor and the matrix transposition operator, respectively. 
Also, $\mathbb{P}$ can be derived as $\mathbb{P}=\mathbb{F}\mathbb{S}$ with $\mathbb{S}$ the second Piola-Kirchhoff stress tensor 
where the constitutive equation \cite{zhang2021sphinxsys} when the material is linear elastic and isotropic is given by
\begin{equation}
\begin{aligned}
\mathbb{S} =\lambda \operatorname{tr}(\mathbb{E}) \mathbb{I}+2 G \mathbb{E},
\end{aligned}
\end{equation}
with $\lambda$ and $G$ denoting the Lame parameter and the shear modulus, respectively.

Then, 
the momentum equation in Eq. \eqref{total Lagrangian SPH governing equatioins} with the weak form \cite{zhang2021sphinxsys} of the SPH particle approximation
can be discretized as 
\begin{equation}
	\label{total Lagrangian SPH governing equatioins discretization}
\rho_i^0 \frac{d\boldsymbol{v}_{i}}{dt}=\sum_j\left(\mathbb{P}_i \mathbf{B}_i^{0^{\mathrm{T}}}+\mathbb{P}_j \mathbf{B}_j^{0^{\mathrm{T}}}\right) 
\cdot \nabla^0 W_{i j} V_j^0,
\end{equation}
where $\nabla^0 W_{i j}=\frac{\partial W\left(\boldsymbol{r}_{i j}^0, h\right)}{\partial r_{i j}^0} \boldsymbol{e}_{i j}^0$ 
and $\mathbf{B}^{0}$ are the kernel gradient and correction matrix in the initial configuration, respectively. 
In addition, the deformation change rate is updated with the strong form of the SPH particle approximation as follows:
\begin{equation}
	\label{deformation change rate}
\frac{d\mathbb{F}}{dt}=\sum_j V_j^0 \left(\boldsymbol{v}_{j}-\boldsymbol{v}_{i}\right) \cdot \nabla^0 W_{i j} \mathbf{B}_i^{0}.
\end{equation}
%
%
%
\section{Numerical results}\label{Numerical results}
In this section, 
a set of numerical simulations including fluid dynamics and elastic solid dynamics
are tested to verify the accuracy and computational efficiency of truncated Laguerre-Gauss kernel (denoted as Laguerre-Gauss) 
and comparison with Wendland kernel \cite{wendland1995piecewise} (denoted as Wendland). 
For clarity, the cut-off radius of both kernels is $2.6dp$.
%
%
\subsection{The Shu-Osher problem}
In this part, 
a one-dimensional Shu-Osher problem is studied to investigate the accuracy of Laguerre-Gauss kernel in the compressible flow. 
The initial condition is given by
\begin{equation}
(\rho,u,v,p)= \begin{cases}(3.857143,2.629369,0,10.3333) &  x\leq1 \\ ($1+0.2sin(5x)$,0,0,1) & \text { otherwise }\end{cases},
\end{equation}
with the domain $x \in [0,10]$ and final time $t=1.8$. 
Also, three spatial resolutions $dp=1/200$, $dp=1/400$ and $dp=1/800$ are applied for the convergence study.
\begin{figure}
	\centering
	\begin{subfigure}[b]{0.49\textwidth}
		\centering
		\includegraphics[trim = 0cm 0cm 0cm 0cm, clip, width=0.9\textwidth]{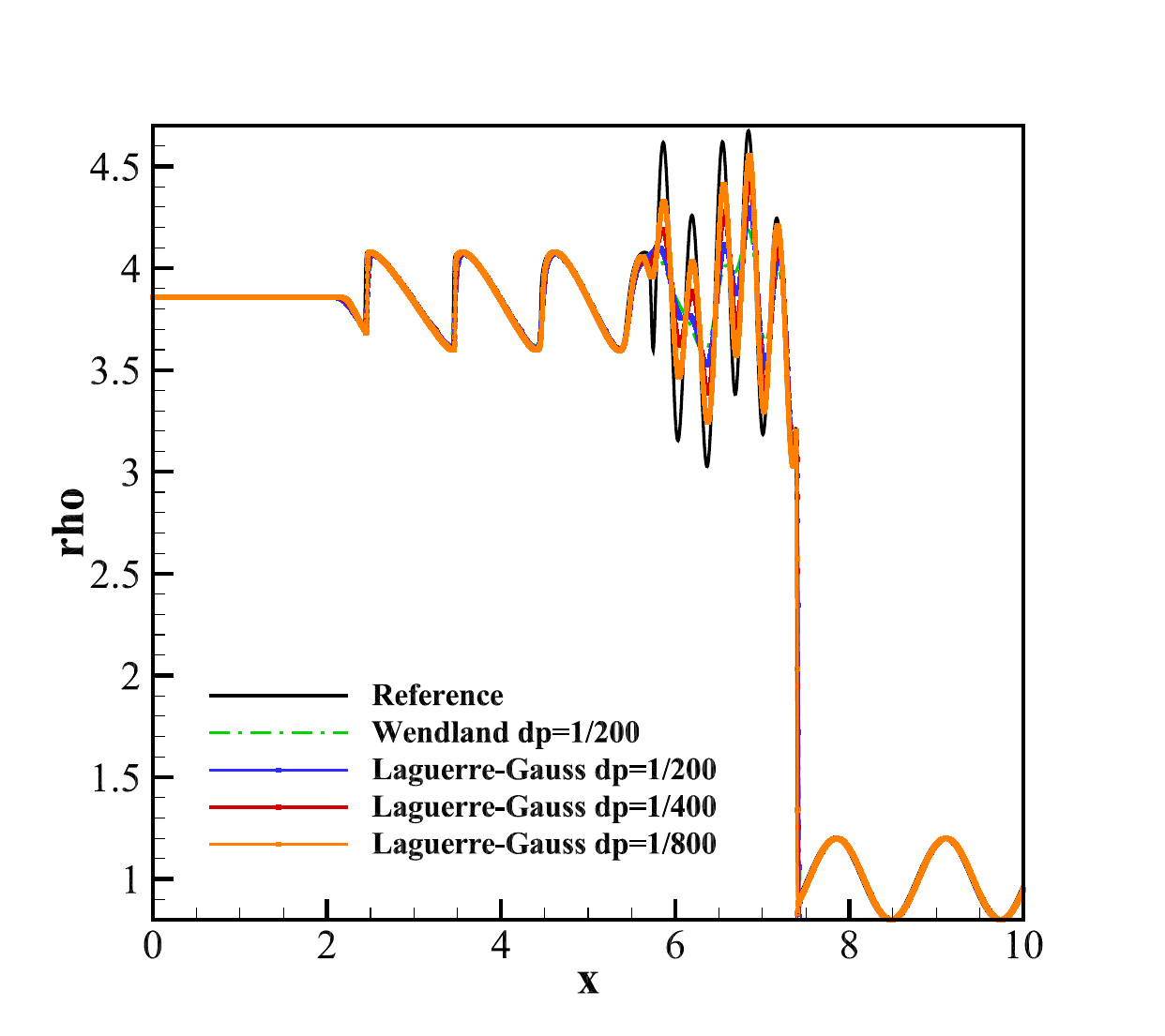}
	\end{subfigure}
		\centering
	\begin{subfigure}[b]{0.49\textwidth}
		\centering
		\includegraphics[trim = 0cm 0cm 0cm 0cm, clip, width=0.9\textwidth]{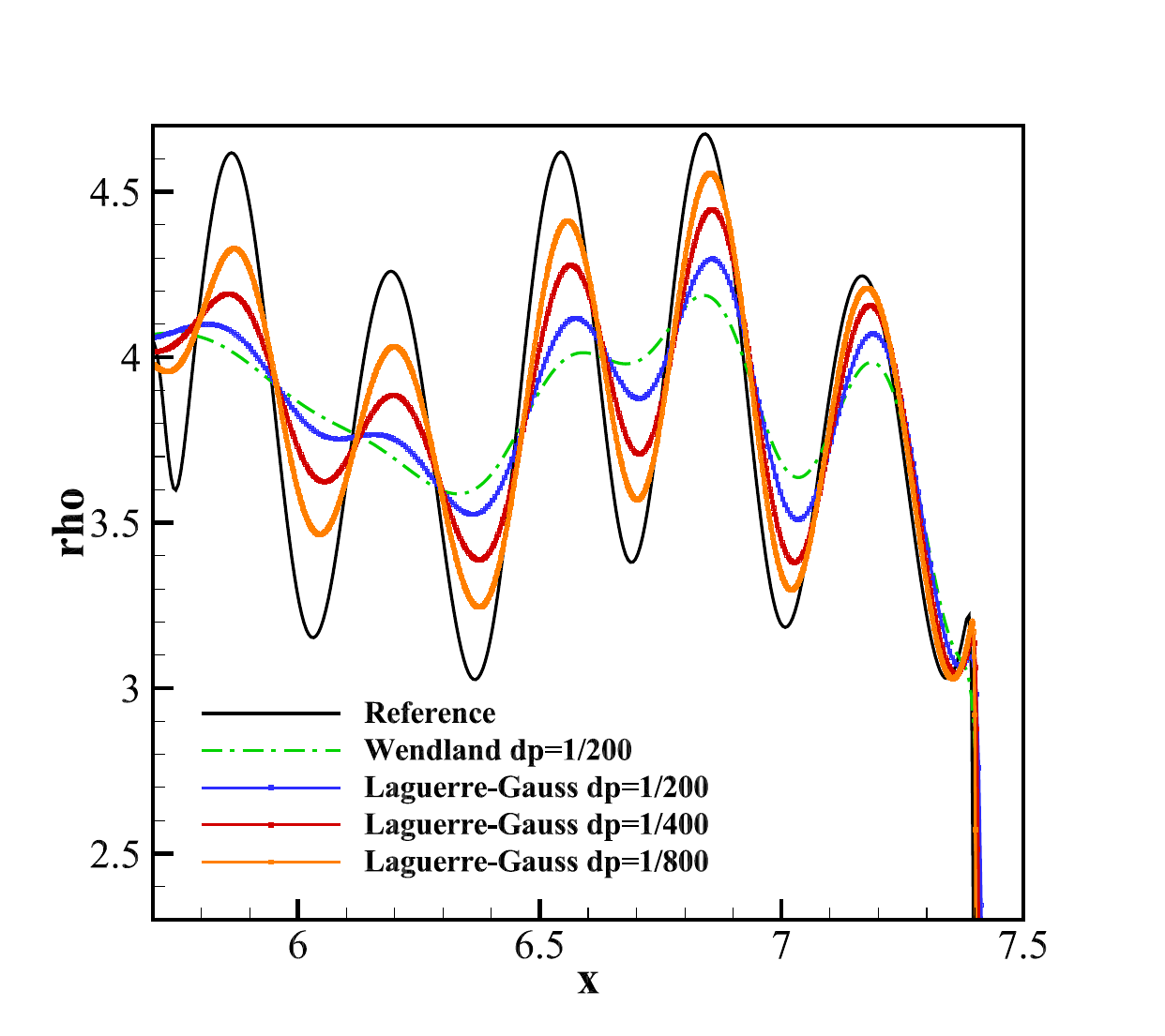}
	\end{subfigure}
	\caption{The Shu-Osher problem: Density profile (left panel) and zoom-in view (right panel) 
	using Wendland and Laguerre-Gauss kernels and their comparisons against the reference data \cite{zhu2020l2}.}
	\label{figs:Shu osher resolution}
\end{figure}

Figure \ref{figs:Shu osher resolution} shows the density and zoom-in view profiles 
using Wendland and Laguerre-Gauss kernels 
and the comparisons with the reference data \cite{zhu2020l2}. 
It can be seen that Laguerre-Gauss kernel can significantly improve the accuracy compared to Wendland kernel at the resolution $dp=1/200$. 
Also,  the results converge rapidly as the resolution increases and there is roughly first-order convergence at the extremities, 
and second-order convergence in regions other than the extremities.
%
%
\subsection{Taylor-Green vortex flow}
In this section, 
we consider the two-dimensional viscous Taylor-Green vortex flow to validate the proposed kernel in the weakly-compressible flow. 
Following Ref. \cite{taylor1937mechanism}, the initial velocity in a unit domain with the periodic boundary conditions in $x-$ and $y-$ directions is given by 
\begin{equation}\label{taylor green initial condition}
	\left\{\begin{array}{l}
	u(x,y,t)=-\exp^{bt}cos(2\pi x)sin(2\pi y) \\
	v(x,y,t)=\exp^{bt}sin(2\pi x)cos(2\pi y) 
	\end{array},\right.
\end{equation}
\begin{figure}
    \centering
    \includegraphics[width=0.5\textwidth]{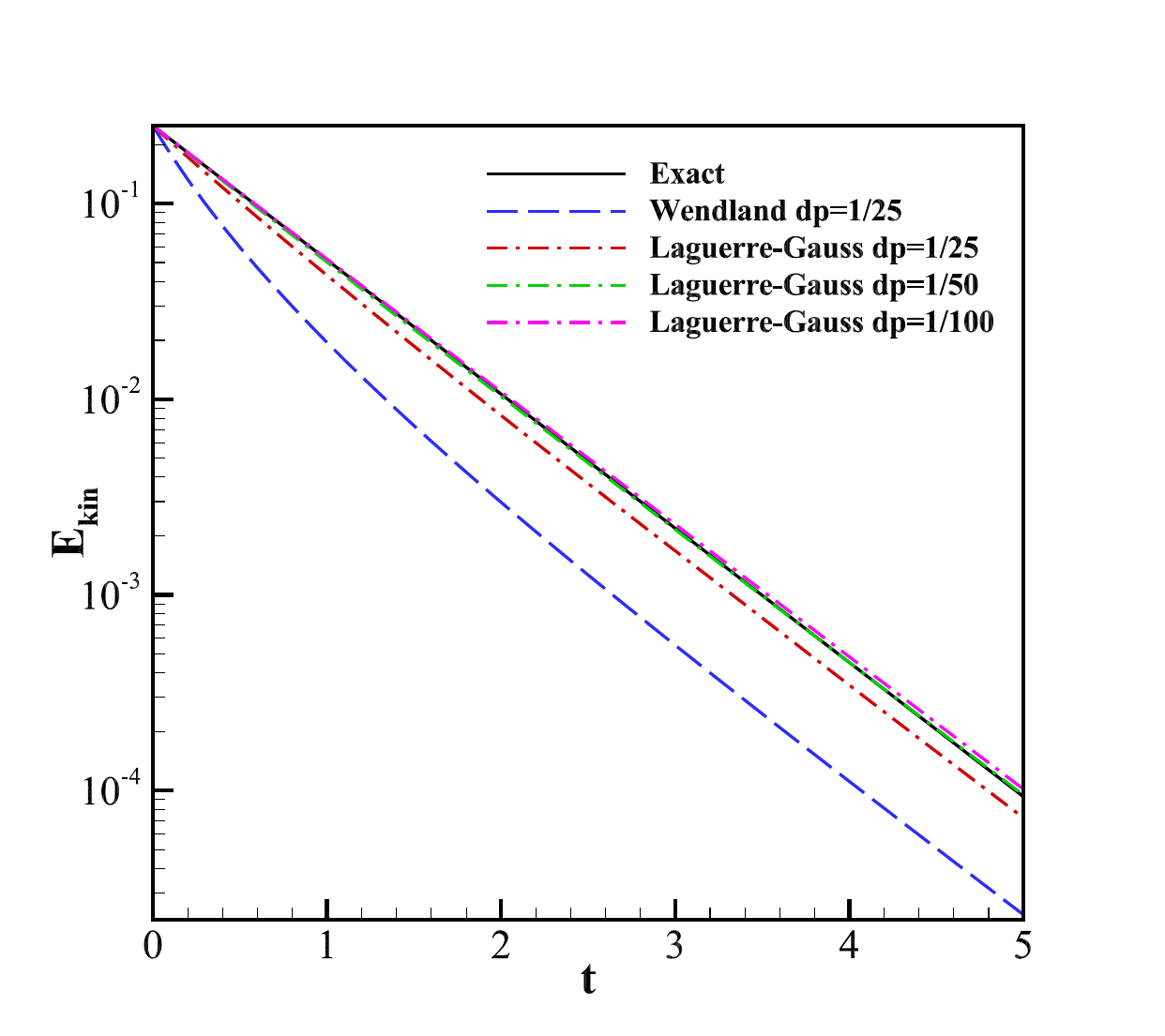}
    \caption{Taylor-Green vortex flow ($Re=100$): Decay of the kinetic energy using Wendland and Laguerre-Gauss kernels 
	and the comparisons with the theoretical solution (denoted as "Exact").}
    \label{Taylor_Green_vortex_energy_decay}
\end{figure}
where the decay rate is $b=-8\pi ^2 /Re$, 
the Reynolds number $Re=100$, the total kinetic energy decay rate $-16\pi ^2 /Re$ and the finial time $t=5$. 
Also, we apply three resolutions including $dp=1/25$, $1/50$ and $1/100$ to test the convergence study.

Figure \ref{Taylor_Green_vortex_energy_decay} portrays the kinetic energy decay 
using Wendland and Laguerre-Gauss kernels and the comparison with the theoretical solution (denoted as "Exact"). 
From the curves of kinetic energy decay, 
the method using Laguerre-Gauss kernel shows clearly improved accuracy by reducing the numerical dissipation more effectively at a resolution of dp=1/25, 
and the results achieve second-order convergence as the resolution increases.
%
%
\subsection{Flow around a cylinder}
In this section, 
flow around a cylinder case containing complex geometries is studied to investigate the versatility of the proposed kernel. 
In order to assess the numerical results quantitatively, the drag and lift coefficients are given as
\begin{equation}\label{eq:wavespeed}
C_{D}=\frac{2F_{D}}{\rho_{\infty}u_{\infty}^2 A},    C_{L}=\frac{2F_{L}}{\rho_{\infty}u_{\infty}^2 A},
\end{equation}
where $F_{D}$ and $F_{L}$ denote the drag and lift force on the cylinder respectively and velocity $u_{\infty}$ 
as well as density $\rho_{\infty}$ in the far-field are set as $1$. 
Also, the Strouhal number $St=fD/u_\infty$ in the unsteady cases 
and the Reynolds numbers $Re=\rho_{\infty}u_{\infty}D/\mu$ is $100$ with the cylinder diameter $D=2$ in the case 
where the far-field boundary conditions are applied in all boundaries and the computational time is $t=300$. 
To alleviate the effects of far-field boundary conditions on the simulations,
the large computational domain size is set as [25D, 15D] where the location of the cylinder centre is (7.5D, 7.5D) and the final time $t=300$
as well as the spatial resolutions are applied as $dp=1/10$, $1/20$ and $1/30$ for the convergence study.
\begin{figure}
	\centering
	\begin{subfigure}[b]{0.49\textwidth}
		\centering
		\includegraphics[trim = 0cm 0cm 0cm 0cm, clip, width=0.9\textwidth]{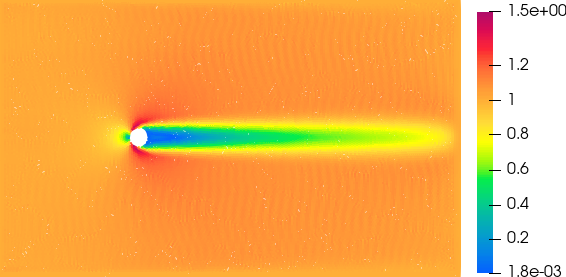}
	\end{subfigure}
		\centering
	\begin{subfigure}[b]{0.49\textwidth}
		\centering
		\includegraphics[trim = 0cm 0cm 0cm 0cm, clip, width=0.9\textwidth]{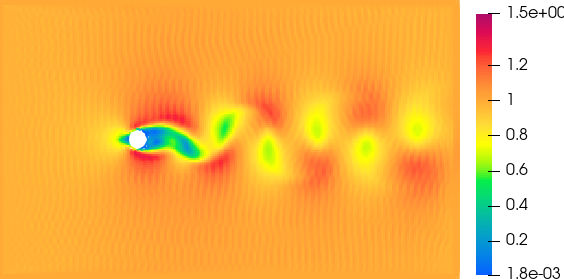}
	\end{subfigure}
	\caption{Flow around a cylinder: Velocity contour ranging from $1.8\times 10^{-3}$ to $1.5$ 
	using Wendland (left panel) and Laguerre-Gauss (right panel) kernels with $Re=100$ under the resolution $dp=1/6$ at $t=300$.}
	\label{flow around cylinder_velocity_contour}
\end{figure}
\begin{figure}
	\centering
	\begin{subfigure}[b]{0.49\textwidth}
		\centering
		\includegraphics[trim = 0cm 0cm 0cm 0cm, clip, width=0.9\textwidth]{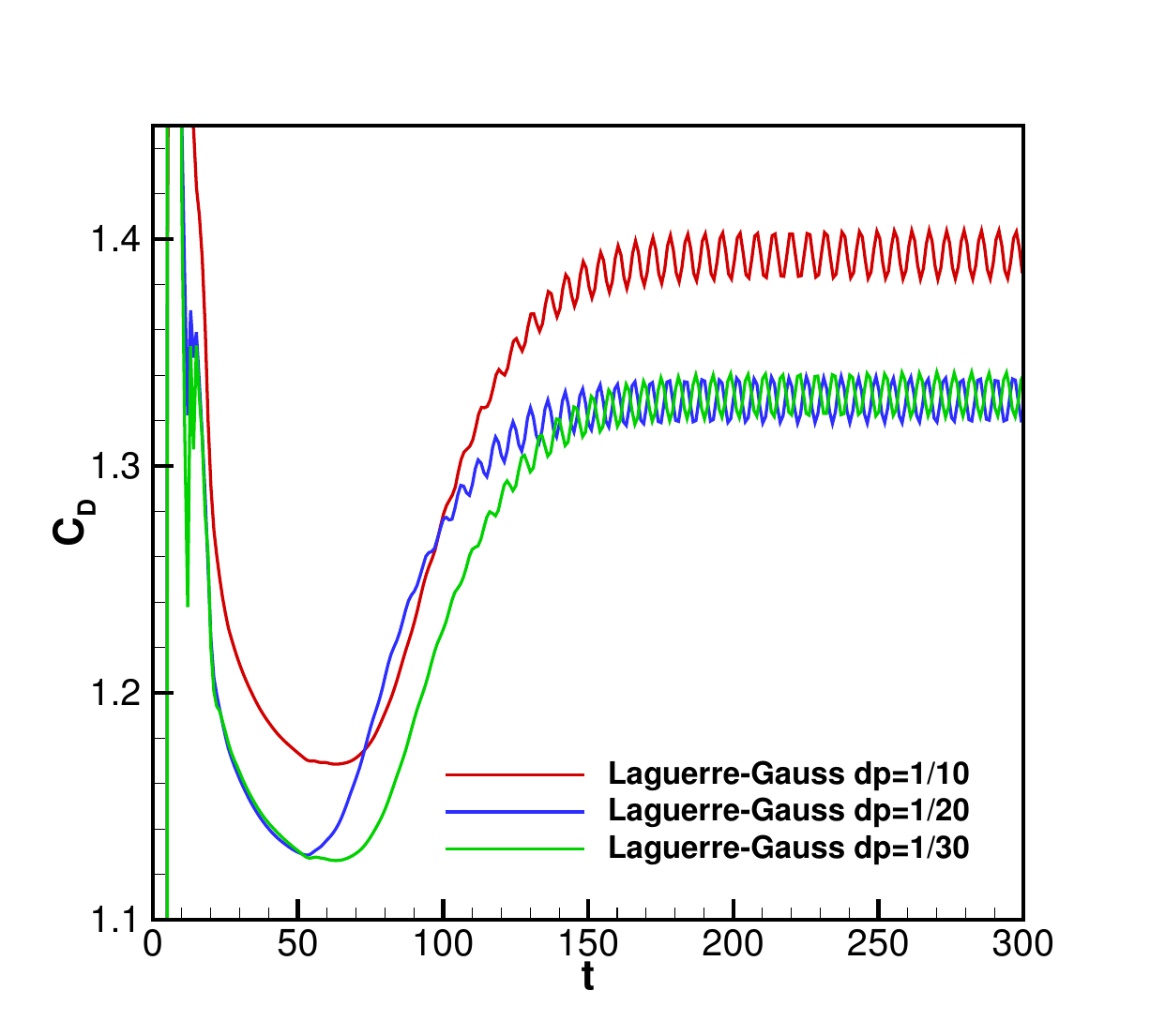}
	\end{subfigure}
		\centering
	\begin{subfigure}[b]{0.49\textwidth}
		\centering
		\includegraphics[trim = 0cm 0cm 0cm 0cm, clip, width=0.9\textwidth]{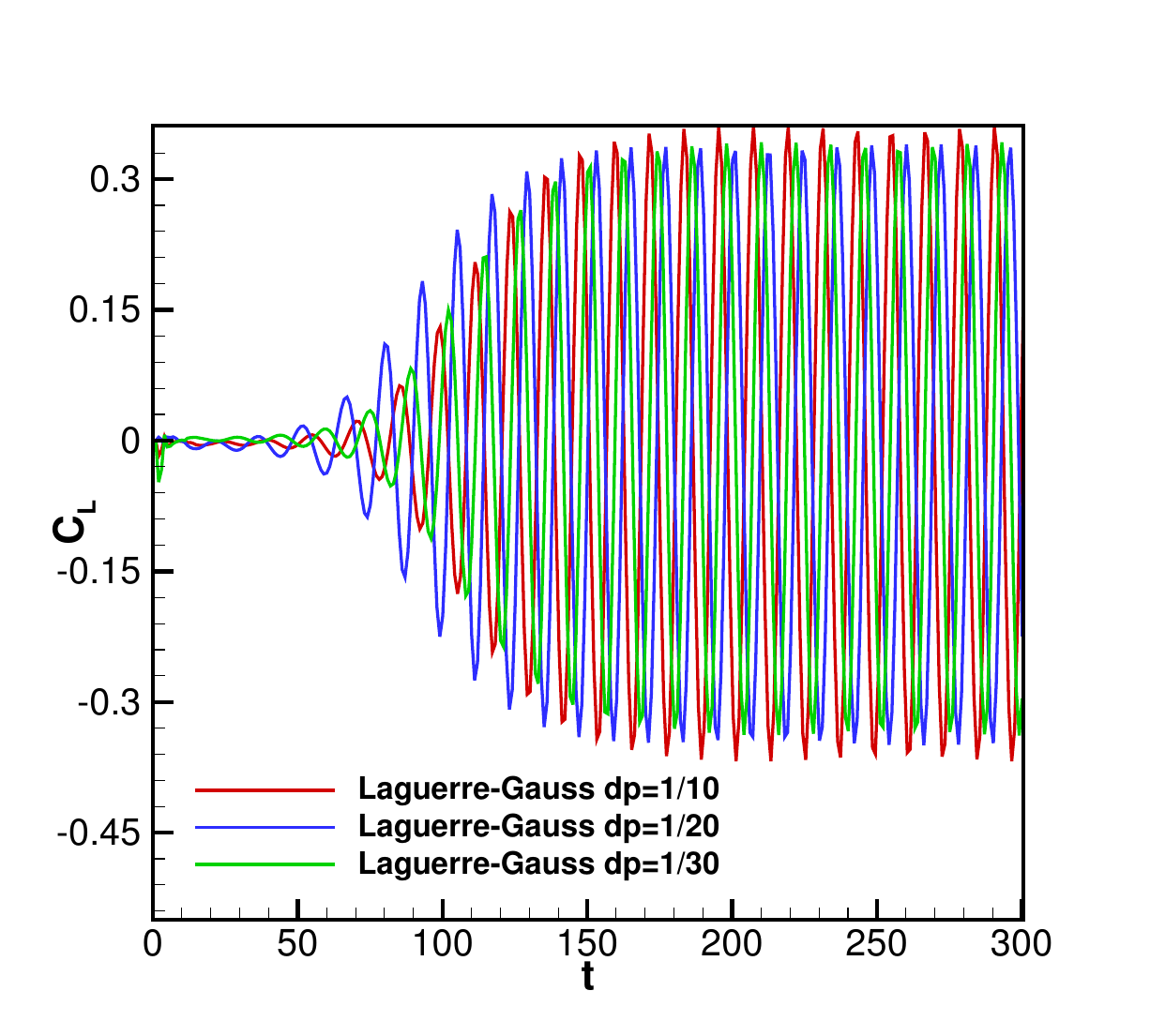}
	\end{subfigure}
	\caption{Flow around a cylinder: Drag (left panel) and lift (right panel) coefficients with the time using
	 Laguerre-Gauss kernel with $Re=100$.}
	\label{flow around cylinder_coefficients}
\end{figure}
\begin{table}
    \caption{Flow around a cylinder: Drag and lift coefficients from different experimental and simulation results at $Re=100$ 
	and the result using Laguerre-Gauss kernel with the resolution $dp=1/30$. }
    \centering
    \begin{tabular}{cccc}
    \hline
    \begin{tabular}[c]{@{}c@{}}Parameters\end{tabular} &
    \begin{tabular}[c]{@{}c@{}}$C_{D}$\end{tabular} &
     \begin{tabular}[c]{@{}c@{}}$C_{L}$\end{tabular} &
     \begin{tabular}[c]{@{}c@{}}$S_{t}$\end{tabular} \\ \hline
    White\cite{white2006viscous} & 1.46 & -  & - \\ \hline
    Khademinejad et al.\cite{khademinezhad2015numerical} & 1.30 $\pm$ - & $\pm$0.292  & 0.152  \\ \hline
    Chiu et al.\cite{chiu2010differentially} & 1.35 $\pm$ 0.012 & $\pm$0.303  & 0.166  \\ \hline
    Le et al.\cite{le2006immersed} & 1.37 $\pm$ 0.009 & $\pm$0.323 & 0.160  \\ \hline
    Brehm et al.\cite{brehm2015locally} & 1.32 $\pm$ 0.010 & $\pm$0.320 & 0.165  \\ \hline
    Liu et al.\cite{liu1998preconditioned} & 1.35 $\pm$ 0.012 & $\pm$0.339 & 0.165  \\ \hline
    Present & 1.33 $\pm$ 0.009 & $\pm$0.340 & 0.169  \\ \hline
    \end{tabular}
    \label{Table_Re=100}
\end{table}

Figure \ref{flow around cylinder_velocity_contour} shows the velocity contour ranging from $1.8\times 10^{-3}$ to $1.5$
using Wendland and Laguerre-Gauss kernels under the Reynolds number $Re=100$ under the resolution $dp=1/6$ at $t=300$. 
It can be seen that the method using Laguerre-Gauss kernel efficiently captures the vortex street 
while that using Wendland kernel fails to present the same physical phenomenon, 
indicating that the proposed kernel is more effective in reducing numerical errors and thus obtaining more accurate results.

We further verify the correctness of the proposed kernel with three resolutions 
by comparing convergent drag and lift coefficients with other experimental and simulation results listed in Table \ref{Table_Re=100} 
where the result of using extended Eulerian SPH with Laguerre-Gauss kernel is denoted as "Present". 
Figure \ref{flow around cylinder_coefficients} presents the drag and lift coefficients 
calculated by Eulerian SPH method using Laguerre-Gauss kernel under three resolutions from $dp=1/10$ to $dp=1/30$ until the final time $t=300$, 
showing that the drag coefficients reach a stable value after a period of fluctuation at the beginning while the lift coefficient oscillates around zero. 
From Figure \ref{flow around cylinder_coefficients} and Table \ref{Table_Re=100}, 
it can be obtained that the results of drag and lift have converged and are in agreement with other references, 
indicating that the results can be considered correct.
In the present study, 
the computations are all performed on an Intel Core i7-10700 2.90 GHz 8-core desktop computer and the CPU wall-clock times
obtained by extended Eulerian SPH method using Wendland and Laguerre-Gauss kernels 
with the spatial resolution $dp=1/10$ are $4604.81s$ and $4818.91s$ in the whole process, respectively, 
indicating that the computational efficiency of using both kernels is comparable.
%
%
\subsection{2D oscillating plate}
In 2D solid dynamics, 
following Refs. \cite{wu2023essentially,gray2001sph,landau1986course}, 
the oscillating plate with the length $L=0.2m$ and the thickness $H=0.02m$ where one edge is fixed and the other edges are free  
is studied to verify the accuracy of Laguerre-Gause kernel. 
In the case,  
the density $\rho_{0}=1000.0 kg/m^{3}$, Young's modulus $E=2$ MPa, physical time $t=2s$ and Poisson's ratio $\nu$ is changeable. 
Also, the initial velocity is perpendicular to the plate strip given by 
\begin{equation}
v_{y}=v_{f}c\frac{f(x)}{f(L)},
\end{equation}
with $v_{f}$ denoting a constant and $f(x)$ written as
\begin{equation}
\begin{split}
f(x)=&(\sin{(kL)}+\sinh{(kL)})(\cos{(kx)}-\cosh{(kx)})-\\
&(\cos{(kL)}+\cosh{(kL)})(\sin{(kx)}-\sinh{(kx)})
\end{split}.
\end{equation}
Here, $k$ is derived by 
\begin{equation}
\cos{(kL)}\cosh{(kL)}=-1,
\end{equation}
with $kL=1.875$. Also, the theoretical frequency $\omega$ is given by 
\begin{equation}
\omega^{2}=\frac{EH^{2}k^{4}}{12\rho (1-\nu^{2})}.
\end{equation}
To keep the results reasonable, 
the largest spatial resolution is set as $H/dp=10$ in the case to discretize the computational domain  
and other two spatial resolutions $H/dp=20$ and $40$ are applied for the convergence study \cite{wu2023essentially,gray2001sph,landau1986course}.
\begin{figure}
    \centering
    \includegraphics[width=1.0\textwidth]{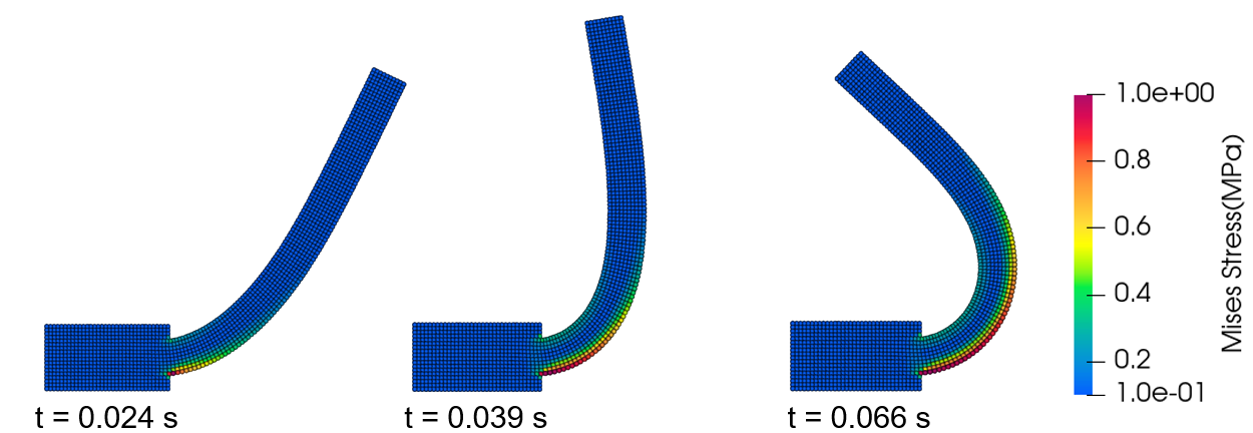}
    \caption{2D oscillating plate: The Mises stress contour ranging from $0.1$ MPa to $1.0$ MPa  
	 using Laguerre-Gauss kernel with $v_{f}=0.15$ and Poisson's ratio $\nu =0.3975$ under the spatial resolution $H/dp=10$.}
    \label{2D_oscillating_beam_contour}
\end{figure}
\begin{figure}
    \centering
    \includegraphics[width=1.0\textwidth]{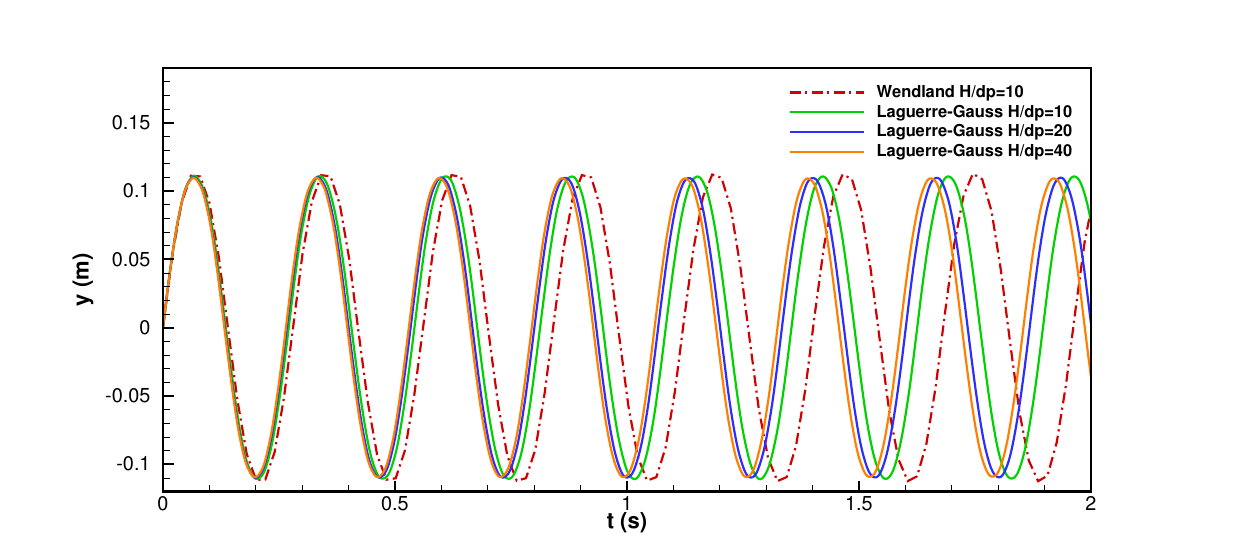}
    \caption{2D oscillating plate: The vertical position y located at the midpoint of the plate strip end \cite{wu2023essentially} 
	using Laguerre-Gauss kernel with the $v_{f}=0.05$ and Poisson's ratio $\nu =0.3975$ 
	under the spatial resolution as $H/dp=10, 20$ and $40$, 
	and the comparison with the result using Wenldand kernel at $H/dp=10$.
	In this case, the theoretical oscillation period $\text{T}_{\text{Theoretical}}=0.254$ 
	and the oscillation periods using truncated Laguerre-Gauss and Wendland kernels are $\text{T}_{\text{LG}}=0.271$ with $\text{Error}_{\text{LG}}= 6.7$\%$ $ 
	and $\text{T}_{\text{W}}=0.282$ with $\text{Error}_{\text{W}}=11.0$\%$ $, respectively.}

    \label{2D_oscillating_convergence_study}
\end{figure}
\begin{table}
    \caption{2D oscillating plate: The oscillation period using Wendland and Laguerre-Gauss kernels with various $v_{f}$ and $\nu $ 
	under the spatical resolution $H/dp=10$ and the comparisons with the theoretical results.}
    \centering
    \begin{tabular}{ccccccc}
    \hline
    \begin{tabular}[c]{@{}c@{}}$v_{f}$\end{tabular} &
    \begin{tabular}[c]{@{}c@{}}$\nu $\end{tabular} &
     \begin{tabular}[c]{@{}c@{}}$\text{T}_{\text{Theoretical}}$\end{tabular} &
     \begin{tabular}[c]{@{}c@{}}$\text{T}_{\text{W}}$\end{tabular} &
     \begin{tabular}[c]{@{}c@{}}$\text{Error}_{\text{W}}$\end{tabular} &
     \begin{tabular}[c]{@{}c@{}}$\text{T}_{\text{LG}}$\end{tabular} &
     \begin{tabular}[c]{@{}c@{}}$\text{Error}_{\text{LG}}$\end{tabular} \\ \hline
    0.05 & 0.22 & 0.27009  & 0.29792 & 10.3$\%$ & 0.28824 & 6.7$\%$ \\ 
   0.1 & 0.22 & 0.27009  & 0.29721 & 10.0$\%$ & 0.29016 & 7.4$\%$ \\ 
   0.05 & 0.3 & 0.26412  & 0.291416 & 10.3$\%$ & 0.28184 & 6.7$\%$ \\ 
   0.1 & 0.3 & 0.26412  & 0.29130 & 10.3$\%$ & 0.28138 & 6.5$\%$ \\ 
    0.05 & 0.4 & 0.25376  & 0.28012 & 10.4$\%$ & 0.27072 & 6.9$\%$ \\ 
   0.1 & 0.4 & 0.25376  & 0.28065 & 10.6$\%$ & 0.27134 & 6.5$\%$ \\ \hline
    \end{tabular}
    \label{oscillation period}
\end{table}

Figure \ref{2D_oscillating_beam_contour} presents the Mises stress contour ranging from $0.1$ MPa to $1.0$ MPa  
using Laguerre-Gauss kernel with the resolution as $H/dp=10$, 
proving that the results obtained by the proposed kernel are smooth.
Also, 
Figure \ref{2D_oscillating_convergence_study} shows the vertical position $y$ at the midpoint of the plate strip end 
using Laguerre-Gauss kernel among the three resolutions and the comparison with that using Wendland kernel with $H/dp=10$. 
It can be seen that as the resolution increases, 
the gap between the different resolutions decreases and therefore the results are convergent and achieve second-order convergence.
Additionally, Laguerre-Gauss kernel demonstrates the ability to attain a more precise oscillation period than Wendland kernel. 
However, 
it is worth noting that the achieved improvement in accuracy is relatively modest when contrasted with previous simulations. 
This is primarily attributed to the fact that the outcomes 
obtained by the latter kernel at lower resolutions are already in close proximity to the theoretical oscillation period.
For quantitative analysis, 
two kernels are applied to calculate the oscillation period $\text{T}$ separately at low resolution $H/dp=10$ to verify the performance. 
Table \ref{oscillation period} lists the oscillation period and errors 
using Wendland (denoted as $\text{T}_\text{W}$ and $\text{Error}_{\text{W}}$, respectively) 
and Laguerre-Gauss (denoted as $\text{T}_{\text{LG}}$ and $\text{Error}_{\text{LG}}$, respectively) kernels 
with different $v_{f}$ and $\nu $ compared with the theoretical oscillation period (denoted as $\text{T}_{\text{Theoretical}}$) under the resolution $H/dp=10$, 
clearly showing that the errors of using Laguerre-Gauss kernel are much smaller than that of using Wendland kernel. 
Furthermore, 
the CPU wall-clock times obtained by total Lagrangian SPH using Wendland and Laguerre-Gauss kernels at the spatial resolution $dp=1/40$ in whole process 
are $537.68s$ and $530.25s$, 
implying that the computational efficiency of using both kernels is comparable.
%
%
\subsection{3D oscillating plate}
In this section, 
we simulate the oscillation of a $3D$ thin plate to validate the robustness and accuracy of the truncated Laguerre-Gauss kernel in $3D$ dimensions. 
Following Refs. \cite{leissa1969vibration,khayyer20213d,khayyer20223d}, 
a square plate is set as the length $L=0.4m$, width $W=0.4m$ and height $H=0.01m$ with the density $\rho_{0}=1000.0kg/m^{3}$, 
Young's modulus $E=100.0$ MPa and Poisson's ratio $\nu=0.3$ and the simulation time is $t=0.1s$. 
Under the condition that the support effect is applied on the centerline of lateral faces, i.e. the corresponding particle is fixed in the $z$-direction, 
the initial velocity $v_{z}$ along the $z$-direction is given as
\begin{equation}
v_{z}(x,y)=\sin{\frac{m\pi x}{L}}\sin{\frac{n\pi y}{W}},
\end{equation}
where $m$ and $n$ control the $x-$ and $y-$ directional vibration modes, respectively. 
Also, the theoretical vibration period of the plate is given by
\begin{equation}
T=\frac{2}{\pi}\left[(\frac{m}{L})^{2}+\frac{n}{W})^{2}\right]^{-1}\sqrt{\frac{\rho H}{D}},
\end{equation}
with the flexural rigidity $D=EH^{3}/\left[12(1-v^{2})\right]$. 
Also, we apply three spatial resolutions $H/dp=5, 7$ and $9$ in the case.
\begin{figure}
    \centering
    \includegraphics[width=1.0\textwidth]{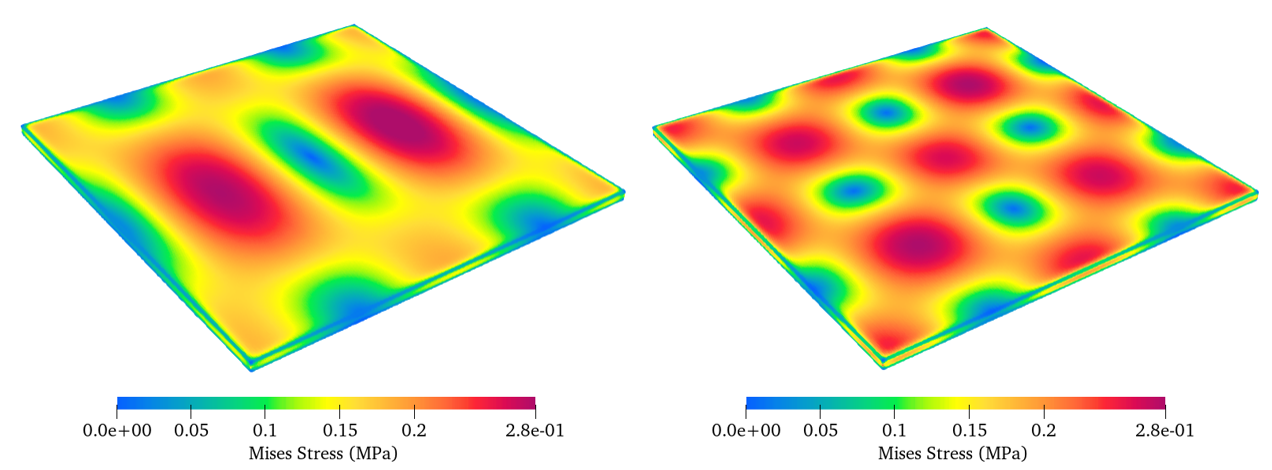}
    \caption{3D oscillating plate: The Mises stress contour ranging from $0$ MPa to $0.28$ MPa 
	 with the vibration mode of $(m,n)=(2,1)$ at $t=0.00279s$ (left panel) and $(m,n)=(2,2)$ at $t=0.00219s$ (right panel) 
	 using Laguerre-Gauss kernel with the spatial resolution $H/dp=9$.}
    \label{3D_oscillating_stress_contour}
\end{figure}
\begin{figure}
	\centering
	\begin{subfigure}[b]{1.0\textwidth}
		\centering
		\includegraphics[trim = 0cm 0cm 0cm 0cm, clip, width=1.0\textwidth]{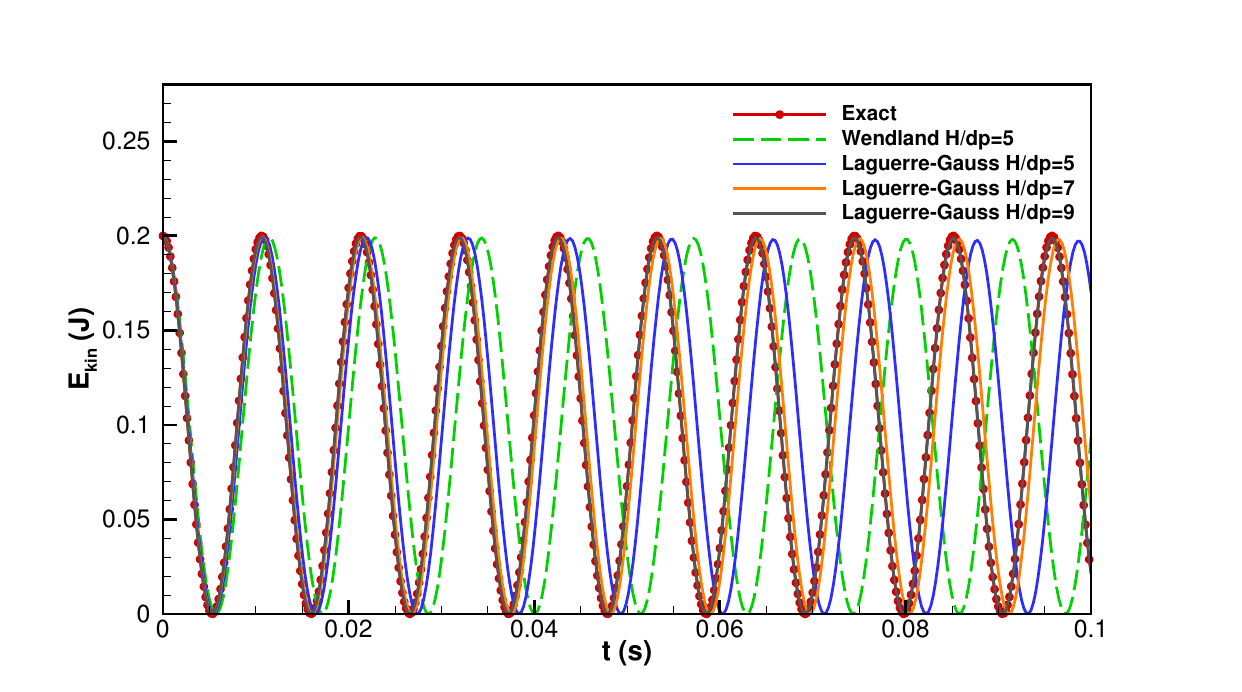}
	\end{subfigure}
		\centering
	\begin{subfigure}[b]{1.0\textwidth}
		\centering
		\includegraphics[trim = 0cm 0cm 0cm 0cm, clip, width=1.0\textwidth]{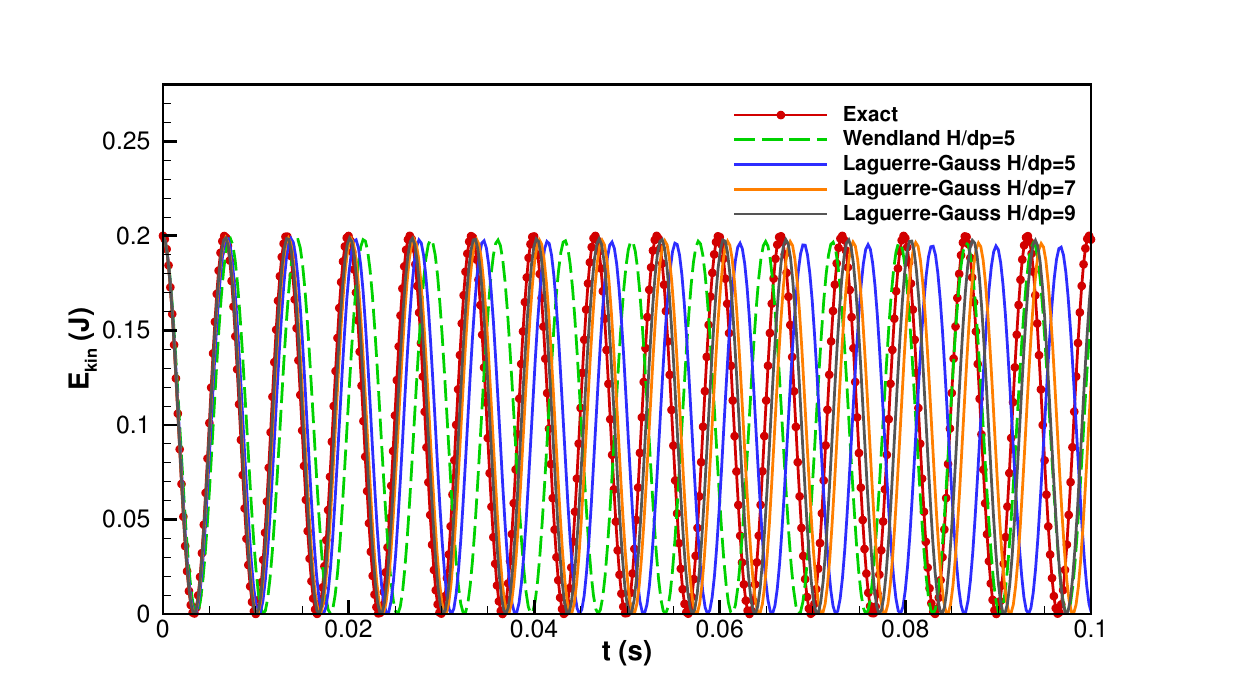}
	\end{subfigure}
	\caption{3D oscillating plate: 
	The oscillation period of the kinetic energy for the vibration mode of $(m,n)=(2,1)$ (upper panel) and $(2,2)$ (bottom panel) 
	using Laguerre-Gauss kernel with the spatial resolution as $H/dp=5, 7$ and $9$, 
	and the comparisons with the results using Wenldand kernel with $H/dp=5$ and the theoretical solutions denoted as "Exact". }
	\label{3D_oscillating_beam_convergence}
\end{figure}

Figure \ref{3D_oscillating_stress_contour} presents the Mises stress contour ranging from $0$ MPa to $0.28$ MPa 
with the vibration mode of $(m,n)=(2,1)$ at $t=0.00279s$ and $(m,n)=(2,2)$ at $t=0.00219s$ 
using Laguerre-Gauss kernel with the spatial resolution $H/dp=9$, 
showing the proposed kernel enable obtain smooth results of Mises stress.
Also, 
Figure \ref{3D_oscillating_beam_convergence} illustrates the oscillation period of the kinetic energy for the vibration mode $(m,n)=(2,1)$ and $(2,2)$ 
using Laguerre-Gauss kernel at three different resolutions $H/dp=5, 7$ and $9$ 
and the comparison with the results using Wenldand kernel with $H/dp=5$ and the theoretical solution, 
implying that Laguerre-Gauss kernel is able to obtain higher accuracy. 
Analogous to the preceding example, 
the enhancement in accuracy using the proposed kernel is relatively marginal due to the same reason.
Besides, 
it is observed that with the increase of spatial resolutions, the results are convergent. 
The wall-clock time for simulation of the vibration mode of $(m,n)=(2,1)$ obtained by total Lagrangian SPH using Wendland and Laguerre-Gauss kernels  
with spatial resolution $H/dp=9$ in whole process are approximately $21910.9s$ and $21906.0s$, respectively, 
showing that the computational efficiency of using Laguerre-Gauss kernel is at the same level as that of using Wendland kernel.
%
%
\subsection{3D Bending column}
We also consider a bending-dominated problem called bending column to investigate the robustness and accuracy of Laguerre-Gauss kernel further. 
Following Refs. \cite{zhang2021integrative,wu2023essentially}, a rubber-like material with a length of $L=6m$, 
a height of $H=1m$ and a square cross-sectional area which is fixed at the bottom is applied in the numerical simulations, 
while the initial velocity conditions are given as $\mathbf{v}_{0}=10(\frac{\sqrt 3}{2},\frac{1}{2},0)^{T}$, density $\rho_{0}=1100kg/m^{3}$, 
Young's modulus $E=17$ MPa and Poisson's ratio $\nu = 0.45$. 
Also, three spatial resolutions including $dp=1/6$, $dp=1/12$ and $dp=1/24$ are applied for the convergence study.
\begin{figure}
    \centering
    \includegraphics[width=1.0\textwidth]{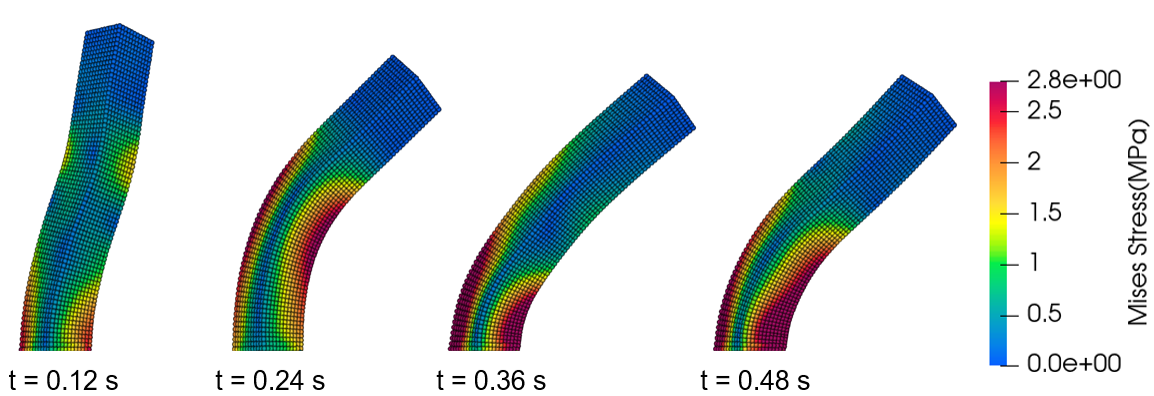}
    \caption{3D Bending column: The Mises stress contour ranging from $0$ MPa to $2.8$ MPa  
	 using Laguerre-Gauss kernel with the spatial resolution $H/dp=12$.}
    \label{3D_Bending_column_contour}
\end{figure}
\begin{figure}
    \centering
    \includegraphics[width=1.0\textwidth]{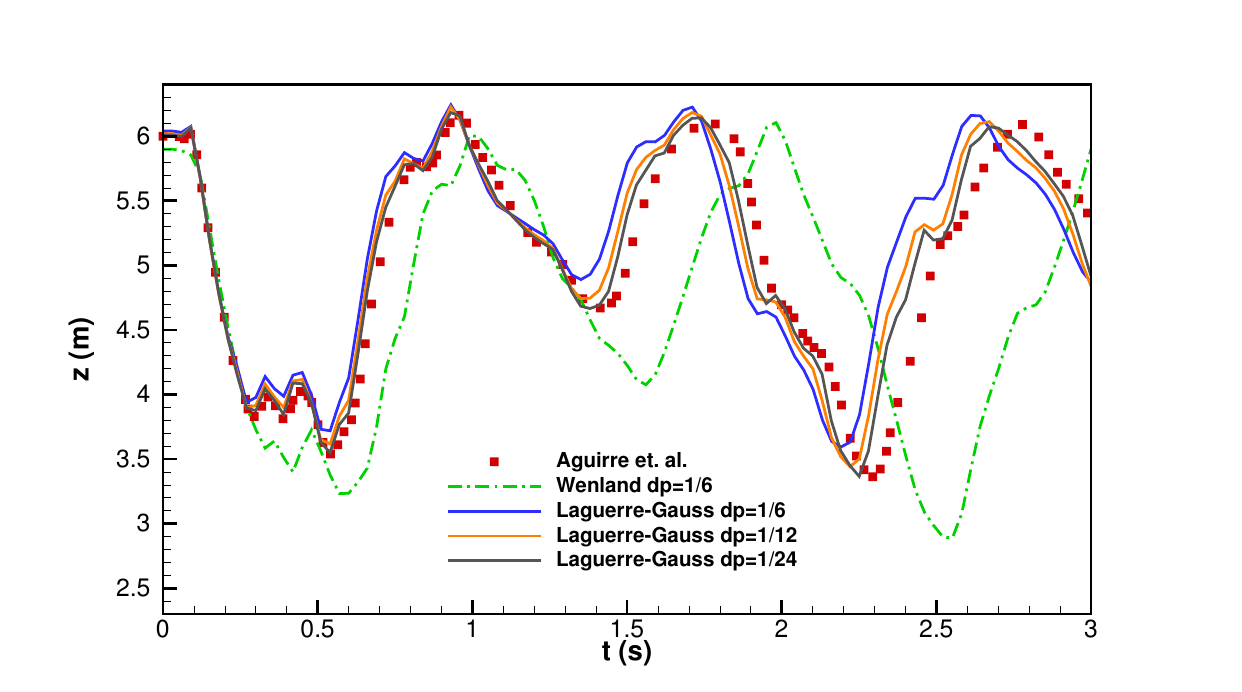}
    \caption{3D Bending column: The vertical position z located at the top of the column \cite{wu2023essentially} 
	using Wendland and Laguerre-Gauss kernels with three different resolutions 
	and the comparisons with the reference data from Aguirre et al. \cite{aguirre2014vertex}.}
    \label{3D Bending column}
\end{figure}

Figure \ref{3D_Bending_column_contour} presents the Mises stress contour ranging from $0$ MPa to $2.8$ MPa 
using Laguerre-Gauss kernel with the spatial resolution $H/dp=12$, 
indicating that the results are smooth by using the proposed kernel.
Besides, 
Figure \ref{3D Bending column} portrays the vertical position $z$ observed at the top of the column \cite{wu2023essentially} 
using Laguerre-Gauss kernel at three different resolutions $dp=1/6, 1/12$ and $1/24$, 
and their comparisons with Wendland kernel at the resolution $dp=1/6$ as well as the result obtained by Aguirre et al. \cite{aguirre2014vertex}, 
showing that Laguerre-Gauss kernel can improve accuracy significantly compared to Wendland kernel as expected 
and the results converge rapidly with the increase of the spatial resolutions.
%
%
\section{Summary and conclusion}\label{Summary and conclusions}
In this paper, 
the error of truncated Laguerre-Gauss kernel is analyzed and 
the results show that the proposed kernel introduces much less truncation error and relaxation residue after particle relaxation 
than Wendland and Laguerre-Wendland kernels, 
indicating its considerably improved accuracy and stability. 
Furthermore,
a set of $2D$ and $3D$ numerical examples for fluid dynamics and solid dynamics are tested and 
have demonstrated that using the proposed kernel enable obtain considerably higher accuracy than Wendland kernel with comparable computational efficiency.
%
%
\bibliographystyle{elsarticle-num}
\bibliography{ref}
\end{document}